\documentclass[journal=nalefd,manuscript=article]{achemso}                          

\setkeys{acs}{maxauthors = 10, etalmode = truncate}                          

\usepackage[version=3]{mhchem} 
\usepackage{graphicx}
\usepackage{color}

\author{Andrey Lyalin}
\email{lyalin.andrey@nims.go.jp}
\affiliation[GREEN]{Global Research Center for Environment and Energy Based on
Nanomaterials Science (GREEN), National Institute for Materials
Science (NIMS), Tsukuba 305-0044, Japan}

\author{Vladimir G. Kuznetsov}
\affiliation[St Petersburg State University]{Faculty of Physics, St Petersburg State University, 
Petrodvoretz, 198504 St Petersburg, Russia}

\author{Akira Nakayama}
\affiliation[Institute for Catalysis]{Institute for Catalysis, Hokkaido University, Sapporo 001-0021, Japan}

\author{Igor V. Abarenkov}
\affiliation[St Petersburg State University]{Faculty of Physics, St Petersburg State University, 
Petrodvoretz, 198504 St Petersburg, Russia}

\author{Ilya I. Tupitsyn}
\affiliation[St Petersburg State University]{Faculty of Physics, St Petersburg State University, 
Petrodvoretz, 198504 St Petersburg, Russia}
\alsoaffiliation[Center for Advanced Studies]{Center for Advanced Studies, Peter the Great St. Petersburg
Polytechnic University, Polytekhnicheskaja 29, 195251 St. Petersburg, Russia}

\author{Igor E. Gabis}
\affiliation[St Petersburg State University]{Faculty of Physics, St Petersburg State University, 
Petrodvoretz, 198504 St Petersburg, Russia}

\author{Kohei Uosaki}
\affiliation[GREEN]{Global Research Center for Environment and Energy Based on
Nanomaterials Science (GREEN), National Institute for Materials
Science (NIMS), Tsukuba 305-0044, Japan}

\author{Tetsuya Taketsugu}
\affiliation[Hokkaido University]{Department of Chemistry, Faculty of Science, Hokkaido University, Sapporo 060-0810, Japan}
\alsoaffiliation[GREEN]{Global Research Center for Environment and Energy Based on
Nanomaterials Science (GREEN), National Institute for Materials
Science (NIMS), Tsukuba 305-0044, Japan}

\title{Lithiation of Silicon Anode based on Soft X-ray Emission Spectroscopy: A Theoretical Study}

\begin{document}
\date{\today}

%\begin{tocentry}
%\begin{center}
%\includegraphics[scale=0.3]{toc.eps}
%\end{center}
%\end{tocentry}

\begin{abstract}
Due to its exceptional lithium storage capacity silicon is considered  
as a promising candidate for anode material in lithium-ion batteries (LIBs). 
In the present work we demonstrate that methods of the soft X-ray emission spectroscopy (SXES) 
can be used as a powerful tool for the comprehensive analysis of the electronic and structural properties of 
lithium silicides \ce{Li_{x}Si} forming in LIB's anode upon Si lithiation. 
On the basis of density functional theory (DFT) and molecular dynamics (MD) simulations 
it is shown that coordination of Si atoms in \ce{Li_{x}Si} decreases with increase in Li concentration 
both for the crystalline and amorphous phases.
In amorphous a-\ce{Li_{x}Si} alloys Si tends 
to cluster forming Si-Si covalent bonds even at the high lithium concentration.
It is demonstrated that the Si-L$_{2,3}$ emission bands of the crystalline and amorphous 
\ce{Li_{x}Si} alloys show different spectral dependencies reflecting the
process of disintegration of Si-Si network into Si clusters and chains of the different sizes
upon Si lithiation. The Si-L$_{2,3}$ emission band of \ce{Li_{x}Si} alloys
become narrower and shifts towards higher energies with an increase in 
Li concentration. The shape of the emission band depends on the relative contribution of the X-ray radiation from the 
Si atoms having different coordination. This feature of  the Si-L$_{2,3}$ spectra of \ce{Li_{x}Si} alloys 
can be used for the detailed analysis of the 
Si lithiation process and LIB's anode structure identification.
\end{abstract}

\maketitle

\section{Introduction}

Lithium-ion batteries (LIBs) are widely used rechargeable power sources for various electronic 
devices.\cite{Armand08,Nitta15} 
The large interest in lithium-ion batteries is stipulated by their compact size, high energy density and operating 
voltage, small memory and self-discharge effects.\cite{Nazri04,Tarascon01,Etacheri11,Sasaki13}. 
However development of the highly efficient batteries progresses slowly, 
due to the lack of suitable electrode materials and electrolytes.\cite{Armand08,Nitta15}
Lithium intercalation of the anode is one of the most important processes in lithium battery that allows it to operate. 
During the charge and discharge cycles lithium diffuses in and out of electrode, inducing changes 
in the anode morphology.\cite{Huggins00,Aifantis07,Chan08,Huang10,Zhao11a} 
The structural disintegrity and cracking results in consequent fading of the capacity 
due to loss of electrical conductance.\cite{Arora98,Spotnitz03,Sarasketa-Zabala14} 
Therefore development of the mechanically stable high-lithium-capacity anode materials based on cheap and abundant elements is an emerging task. 

Silicon is one of the very promising candidates for anode materials in LIBs 
because it exhibits more than order of magnitude greater theoretical Li capacity (4200 mAh/g for \ce{Li_{4.4}Si}) 
than conventional graphite anodes (372 mAh/g).\cite{Sharma76,Boukamp81,Winter99} 
Moreover, Si is cheaper and more abundant than graphite. However, the high capacity of Si to host Li atoms 
results in a large volume expansion of about 400\%, amorphization of Si and crumbling of the electrode.\cite{Obrovacz04}
To overcome this problem one should have detailed understanding of mechanisms of structural transformations 
in Si electrodes upon lithiation/delithiation processes.\cite{Chiang16}
Considerable efforts have been made to elucidate the origin of the degradation performance of Si anodes in LIBs 
with the aim to improve their stability.\cite{Chiang16,Soni11,Xie10,Balke10,Yoshimura07,Ding09} 
The lithiation process of crystalline Si (c-Si) has been intensively investigated  both experimentally 
and theoretically.\cite{Kubota08,Zhang10,Zhao11a,Chon11,Kim11,Liu12,McDowell12,Chan12,Pharr12,Jung12,Morris13,Cubuk14} 

Recent theoretical studies have been mainly focused on the modelling of lithium insertion and diffusion 
in c-Si,\cite{Wan10,Peng10,Chan12} 
search for the stable crystalline \ce{Li_{x}Si} phases\cite{Kubota08,Chevrier09,Chevrier10,Zhao11a,Morris14,Valencia-Jaime16}
and investigation of the structural and dynamic properties of amorphous \ce{Li_{x}Si} alloys.\cite{Chevrier10a,Huang11,Kim11,Chiang16}
It has been demonstrated that several  lithium silicide crystalline phases,
\ce{LiSi},\cite{Stearns03,Kubota08}                  
\ce{Li_{12}Si_{7}},\cite{Nesper86,Leuken96}          
\ce{Li_{7}Si_{3}},\cite{Boehm84}                     
\ce{Li_{13}Si_{4}},\cite{Zeilinger13}             
\ce{Li_{15}Si_{4}},\cite{Zeilinger13a,Kubota07}      
\ce{Li_{17}Si_{4}},\cite{Zeilinger13b} and               
\ce{Li_{22}Si_{5}}\cite{Nesper87}
can be stable.
These crystalline structures can be formed during high-temperature lithiation, while 
room temperature lithiation of c-Si often results in formation of amorphous lithium silicides a-\ce{Li_{x}Si}.\cite{Limthongkul03}                 
It was suggested that the lack of formation  of crystalline c-\ce{Li_{x}Si} alloys at room temperature is 
most likely due to kinetic constraints, which means that c-Si lithiation at the room temperature is a nonequilibrium 
process.\cite{Chan12} After the first cycle of lithiation and delithiation c-Si becomes amorphous.\cite{Liu12} 
However it has been found that lithiation of c-Si is a multi-phase process, where c-Si is 
lithiated layer by layer.\cite{Zhao11a,Chan12,Liu12,Cubuk13,Cubuk14,Seidlhofer16}
Operando neutron reflectivity analysis has demonstrated that 
lithiation starts with the formation of a lithium enrichment zone during the first charge step.\cite{Seidlhofer16} 
The Li enriched area of \ce{Li_{x}Si} can be divided into a highly lithiated zone at the surface with concentration x $\sim$ 2.5  
and a much less lithiated growth region with  x $\sim$ 0.1 formed deep into the crystal. 
The  thickness of the highly lithiated zone is the same for the first and second cycle, 
whereas the thickness of the less lithiated zone is larger for the second lithiation.\cite{Seidlhofer16}
It has also been found that lithiated Si is separated from pristine c-Si by a sharp reaction front with thickness of about 1 nm, 
which moves into c-Si as the reaction progresses.\cite{Chon11,Liu12,Cubuk14} 
It is interesting that in the case of  amorphous Si recent experiments also indicate 
the presence of the boundary between lithiated and pristine a-Si 
in spite of the absence of the crystalline order in a-Si.\cite{Wang13,McDowell13}
It has been suggested that as the reaction front progresses into pure a-Si, 
the lithiated Si has relatively constant Li concentration x in a-\ce{Li_{x}Si} of x $\sim$ 2.5.\cite{Wang13}
After initial lithiation to x $\sim$ 2.5, a second step of the reaction occurs, where 
Li concentration increases from x $\sim$ 2.5 to x $\sim$ 3.75.\cite{Wang13}
Interestingly, several groups reported\cite{Choi14,Zeng13,Son11,Liu12a,Misra12,Gu13,Liu11} that crystalline c-\ce{Li_{15}Si_{4}} 
can be formed through an amorphous \ce{Li_{x}Si} phase, however 
in other works  only formation of the
amorphous \ce{Li_{x}Si} phases was observed.\cite{Chan09,Long11,Schroder12,Liu12,Kang14,Pereira-Nabais13}

Detailed understanding  of the mechanisms of such multi-phase electrochemical lithiation processes is of 
extraordinary importance for the development of the stable electrodes for high-performance LIBs.
Such processes have been extensively studied experimentally by  
transmission electron microscope (TEM),\cite{Chon11,Liu12,Wang13,McDowell13}
X-ray diffraction (XRD),\cite{Limthongkul03,Li07}
nuclear magnetic resonance (NMR),\cite{Key09,Key11}
electron energy loss spectroscopy (EELS),\cite{Kang09b} 
neutron reflectometry (NR)\cite{DeCaluwe15,Veith15,Seidlhofer16}  methods. 
However the results of experimental studies on the structural properties of electrochemically 
lithiated Si are still very controversial,
do not provide complete understanding of the silicon lithiation process
and do not give information about chemical state and electronic properties of the \ce{Li_{x}Si} based anode 
material in LIBs.

Recently, the structure, composition and electronic state of electrochemically lithiated Si(111) 
have been studied by methods of soft X-ray spectroscopy (SXES) combined with the X-ray 
diffraction with synchrotron radiation.\cite{Aoki16}
It has been reported that three different phases of electrochemically lithiated Si are likely formed on the Si(111) substrate:
(i) a single-crystalline c-\ce{Li_{15}Si_{4}} alloy phase, 
(ii) an amorphous phase of a-\ce{Li_{15}Si_{4}} and/or a-\ce{Li_{13}Si_{4}}, and 
(iii) a mixed phase of a-\ce{Li_{15}Si_{4}} and/or a-\ce{Li_{13}Si_{4}}  (52 \%)
with the crystalline c-Si (48 \%).\cite{Aoki16} 
However the detailed interpretation of the experimental results has been hindered 
due to the absence of the theoretical data 
on the formation mechanisms of the  Si-L$_{2,3}$ emission of \ce{Li_{x}Si} alloys. 

In the present work we  perform theoretical analysis of the  mechanisms of formation of the soft X-ray Si-L$_{2,3}$ 
emission of crystalline and amorphous \ce{Li_{x}Si} alloys. 
On the base of comparison of results of our calculations with the available 
experimental data we demonstrate that methods of SXES can be used as a powerful tool for the comprehensive 
analysis of the electronic and structural properties of crystalline and amorphous \ce{Li_{x}Si} alloys in LIBs.
In particular it is shown that the energy position and shape of the Si-L$_{2,3}$ band provides
information about disintegration of the Si network
into Si clusters of the different sizes upon Si lithiation, as well as 
chemical structure and composition of \ce{Li_{x}Si} alloys. 
Therefore SXES methods can be used as a powerful tool for investigation of the
lithiation process of Si and multi-phase transitions in the 
crystalline and amorphous structures.

\section{Methods}

The calculations reported herein were performed using the density-functional theory (DFT) method  
in a  plane wave (PW) basis set\cite{Payne92} as implemented in 
the pseudopotential-based CASTEP code.\cite{Segall02,Clark05}  
We used  the generalized gradient approximation (GGA) with the parametrization 
of Perdew-Burke-Ernzerhof (PBE)\cite{Perdew96} for the exchange-correlation functional  
and the ultrasoft pseudopotentials (USPs)\cite{Vanderbilt90}  with two projectors for each angular 
momentum  to describe the electron-ion interactions. The cut-off energy of 280 eV has been used.  
In the used UCPs  the  1$s^2$2$s^1$ electrons  of  Li atom  and  the 3$s^2$3$p^2$ electrons of Si atom 
have been treated as valence electrons.
The structures of the crystalline Si (c-Si)\cite{Wyckoff63} and four stable lithium silicide crystalline phases 
with the different concentration x of Li atoms, 
c-\ce{LiSi}\cite{Evers97,Stearns03} (x=1),  
c-\ce{Li_{12}Si_{7}}\cite{Leuken96,Wu07} (x=1.71),  
c-\ce{Li_{13}Si_{4}}\cite{Zeilinger13} (x=3.25), 
and  c-\ce{Li_{15}Si_{4}}\cite{Zeilinger13a} (x=3.75) 
have been fully optimized and relaxed.
The robust Broyden-Fletcher-Goldfarb-Shanno (BFGS) optimizer with line search has been used for 
optimization of cell parameters and the all-bands conjugate-gradient minimizer has been used to determine 
the relaxed atomic positions.
The geometry optimization has been  performed until the energy difference per atom, the forces on the atoms and all the stress components 
do not exceed the values 1$\times$$10^{-6}$ eV/atom, 2$\times$$10^{-3}$ eV/{\AA} and 4$\times$$10^{-3}$  GPa, respectively. 
The convergence of the self-consistent energy was achieved with a tolerance of 5$\times$$10^{-7}$ eV/atom.

The Monkhorst-Pack\cite{Monkhorst-Pack} k-point meshes (12$\times$12$\times$12), (5$\times$5$\times$8), (5$\times$2$\times$3), 
(6$\times$3$\times$11), and 
(5$\times$5$\times$5) have been used for Brillouin zone sampling of 
c-\ce{Si}, c-\ce{LiSi}, c-\ce{Li_{12}Si_{7}}, c-\ce{Li_{13}Si_{4}}, and  c-\ce{Li_{15}Si_{4}} structures, respectively. 
The used k-point meshes
were chosen in such a way that
the respective  maximum k-point spacings
0.0305 \AA$^{-1}$ for c-\ce{Si},\,\ 
0.0218 \AA$^{-1}$ for c-\ce{LiSi},\,\ 
0.0233 \AA$^{-1}$ for c-\ce{Li_{12}Si_{7}},\,\ 
0.022 \AA$^{-1}$ for c-\ce{Li_{13}Si_{4}},\,\ 
and \,\ 0.0188~\AA$^{-1}$ for c-\ce{Li_{15}Si_{4}},
is  be approximately the same for different structures, 
which corresponds roughly the same accuracy of calculations.
The calculated lattice constants proved to be in  an excellent agreement with the corresponding experimental values, as
shown in Table \ref{tbl:lattice}. Slight overestimation of the lattice constants with respect to the experiment 
is a general feature of PBE type of density functionals.\cite{Haas09}

\begin{table}
\caption{Optimized lattice constants of the considered crystalline Li--Si structures and the corresponding experimental values (in parentheses).}
\label{tbl:lattice}
\begin{tabular}{lll}
\hline
Structure           & Space group & Lattice constants (\AA)  \\
\hline
\ce{Si}             & Cm(8)       & a = 5.464 (5.431)\cite{Wyckoff63}                                         \\
\ce{LiSi}           & I41/a(88)   & a = 9.365 (9.353), c = 5.761 (5.743)\cite{Evers97}                        \\        
\ce{Li_{12}Si_{7}}  & Pnma(62)    & a = 8.574 (8.596), b =  19.709 (19.775), c = 14.361 (14.319)\cite{Wu07}   \\
\ce{Li_{13}Si_{4}}  & Pbam(55)    & a = 7.958 (7.949), b = 15.158  (15.125), c = 4.451 (4.466)\cite{Zeilinger13}   \\
\ce{Li_{15}Si_{4}}  & I$\bar{\rm 4}$3d(220)  & a = 10.654 (10.632)\cite{Zeilinger13a}                         \\
\hline 
\end{tabular}
\end{table}

For further modeling of the amorphous structures we constructed 2x2x2 supercell for \ce{Si} and 1x1x2 supercell 
for \ce{LiSi} and \ce{Li_{13}Si4} phases in order to maintain the total number of atoms in the cell not less than 64. 
More specifically, the numbers of Li and Si atoms in the simulation cell are set to 
(Li:Si) = (0:64), (32:32), (96:56), (52:16) and (60:16) for \ce{Si}, \ce{LiSi}, \ce{Li_{12}Si7}, 
\ce{Li_{13}Si4} and \ce{Li_{15}Si4} phases, respectively.  
The amorphous structures of Li-Si alloys were generated by the first-principles molecular dynamics simulations via the melt-and-quench scheme, 
where we used the CP2K package\cite{VandeVondele05} with the mixed Gaussian and plane-waves (GPW) approach. 
The PBE functional was employed as the exchange and correlation potential.  
The double-$\zeta$ valence plus polarization (DZVP) basis sets of the MOLOPT type\cite{VandeVondele07} were 
used to represent the atomic orbitals of the out-core  
electrons (3 and 4 electrons for Li and Si, respectively) in conjunction with 
the norm-conserving Goedecker-Teter-Hutter pseudopotentials.\cite{Goedecker96,Hartwigsen98}  
The energy cutoff of 400 Ry was used for the auxiliary plane wave expansion of the density.  
Only the $\Gamma$-point was considered in a supercell approach. 
For \ce{LiSi}, \ce{Li_{12}Si7},  \ce{Li_{13}Si4}, \ce{Li_{15}Si4} alloys, starting from the crystal structure, 
the system was first melted at 1500 K under 1 atm in the NPT ensemble for 20 ps, where the temperature 
and pressure was controlled by thermostats and barostats, respectively, using an isotropic cell with variable cell lengths.  
The time step was set to 1.0 fs.  Then, the system was annealed gradually to 300 K during a run of 5 ps, 
followed by 5 ps of equilibration at 300 K. 
In the case of  a-\ce{Li_{13}Si4} and  a-\ce{Li_{15}Si4} four independent samples were used for further calculations
to have good statistics. For modelling amorphous Si, the melting temperature was set to 2500 K 
with longer simulation time of 100 ps due to the 
high melting point of Si, and then the system was annealed to 300 K during a run of 30 ps, followed by 5 ps of equilibration at 300 K.

Neglecting the finite width of the core level, the intensity $I(E)$ of an X-ray emission spectrum (XES) is given by the following expression:\cite{Ovcharenko11} 
\begin{equation}
I_0(E) = E \frac{1}{N}  \displaystyle\sum_{n,{\bf k}} P_n ({\bf k})  \delta(E - E_n({\bf k}) + E_c),
\label{eq:1}
\end{equation}
\noindent where the summation over $\bf k$ is performed over occupied states of the Brillouin zone, 
$E = \hbar \omega$ is the energy of the X-ray photon,
$E_n({\bf k})$ is the one-electron energy of the $n$th valence band,  
$E_c$ is the energy of the core level,
$N$ is the number of points in the  Brillouin zone, 
and $P_n({\bf k})$ is the probability of transition from $n$th valence band to the core level $c$ per unit time.

Atomic system of units $\hbar = e = m = 1$ are used throughout unless specified otherwise. 
 
In the one-electron and dipole approximation, the probability $P_n({\bf k})$  of the radiative electron transition 
is determined by the formula
\begin{equation}
P_n({\bf k}) =  \frac{4}{3} \left(\frac{\omega_{nc}({\bf k})}{c}\right)^3 \frac{1}{2l_c + 1}  \displaystyle\sum_{m_c,\alpha}  \left|\left<  \psi_{n{\bf k}}  \middle| \ r_\alpha\middle| \phi_c \right>\right|^2,
\label{eq:2}
\end{equation}
\noindent where $\omega_{nc}({\bf k})$ is the transition frequency, $l_c$ is the orbital quantum number of the core vacancy and 
$c$ is the speed of light,
\noindent  $\psi_{n{\bf k}}({\bf r})$ and  $\phi_c({\bf r})$ are the wave functions of the $n$th band and the hole, respectively, 
$m_c$ is the quantum number of the projection of the core hole angular momentum and index $\alpha= -1, 0, 1$ enumerates cyclic components 
of the position operator $\ r_\alpha $. 
%$\displaystyle\frac{1}{2l_c + 1}  \displaystyle\sum_{m_c}$ corresponds to averaging over degenerate core states.

In the present work the crystal wave functions $\tilde \psi_{n{\bf k}}({\bf r})$  are calculated 
using the plane wave basis and the ultrasoft pseudopotential scheme implemented in  
the CASTEP software package.\cite{Segall02,Clark05}
Such single-electron pseudo (PS) wave functions are smoothed in the atomic core regions
and can not be used directly for
for calculating the 
probability of the radiative electron transition given by Eq. \ref{eq:2},
because it is determined by the behavior of the wave functions near the nucleus.
Therefore in order to calculate the dipole matrix element in Eq. \ref{eq:2} the 
all-electron (AE) orbitals have been recovered from  PS wave functions
using the projected augmented wave (PAW) reconstruction method\cite{Bloechl94} 
implemented in CASTEP code.\cite{Segall02,Clark05} 
Different methods of recovering of AE orbitals along with the PAW method are 
described in details in the previous works.\cite{Abarenkov01,Ovcharenko11} 
According to the PAW approach, an AE wave function 
$\psi_{n{\bf k}}({\bf r})$ can be recovered from
the corresponding PS wave function $\tilde \psi_{n{\bf k}}({\bf r})$ by a linear transformation
\begin{equation}
\psi_{n{\bf k}}({\bf r}) \,=\, \tilde \psi_{n{\bf k}}({\bf r}) \,+\,  \displaystyle\sum_{i} (\phi_{i}  -  \tilde \phi_{i}) \langle \tilde p_{i} \mid \tilde \psi_{n{\bf k}}({\bf r}) \rangle \,,
\label{eq:3}
\end{equation}
\noindent where  $\phi_{i}$ and $\tilde \phi_{i}$ are AE and PS partial waves, respectively, and $\tilde p_{i}$ 
are the PAW projector functions\cite{Bloechl94} localized within the augmentation region and forming a basis set 
dual to $\tilde \phi_{i}$, i.e. $\langle \tilde p_{i} \mid \tilde \phi_{i} \rangle \,= \, \delta_{i,j}.$
%\begin{equation}
%\langle \tilde p_{i} \mid \tilde \phi_{i} \rangle \,= \, \delta_{i,j}.
%\label{eq:4}
%\end{equation}
Using this transformation the dipole matrix element can be written as follows
\begin{equation}
\left<  \psi_{n{\bf k}}  \middle| \ r_\alpha \middle| \phi_c \right> \,=\, \left< \tilde  \psi_{n{\bf k}}  \middle| \ r_\alpha\middle| \phi_c \right>
\,+\, \displaystyle\sum_{i} \left(\left< \phi_{i}  \middle| \ r_\alpha\middle| \phi_c \right> \,-\, 
\left< \tilde  \phi_{i}  \middle| \ r_\alpha\middle| \phi_c \right>\right) \langle \tilde p_{i} \mid \tilde \psi_{n{\bf k}}({\bf r}) \rangle \,. 
\label{eq:5}
\end{equation}
To calculate the dipole matrix elements and intensity of XES we used the on-the-fly generated (OTFG)
ultrasoft pseudopotentials developed by Pickard as implemented in CASTEP code.\cite{Segall02,Clark05} 

Note that the OTFG pseudopotentials were originally developed  in order to account for the 
core hole effect in the model of supercell by generating the  pseudopotential of "excited" atom 
with the core hole.\cite{Pickard97,Gao08,Gao09}. 
However, in semiconductors and especially in metals the core hole can be effectively 
screened by valence electrons, which allows to calculate XES without accounting for the 
core hole effect on XES. Indeed, our qualitative estimates, given in the Supporting Information, 
show that in the case of "ƒ-Si the core hole is screened before the spontaneous X-ray emission 
transition occurs. In the case of lithium silicides which possess metallic properties 
such approximation should work even better. 
Therefore, in the present work we do not take into account the core hole effect on XES.
In this case  the generated OTFG pseudopotential of Si atom differs from the usual ultrasoft pseudopotential 
with two projectors for each of  s-  and  p-  channels only by addition of two gamma-projectors into local d- channel 
during an automated generating with the help of Materials Studio graphical interface.\cite{MatStudio70}     

The CASTEP calculations do not take into account the spin-orbit interaction. 
Therefore, it is not possible to distinguish 
the calculated XES between $L_2$ and $L_3$ emission bands, which are formed  as a result of the electron transitions 
from the valence bands to the 2$p_{1/2}$ and 2$p_{3/2}$ core hole states, respectively. 
Therefore, to compare correctly the calculated $L_{2,3}$ XES with  the experimental ones 
we have used a simple model procedure 
generating an approximate $L_3$ spectrum by shifting the calculated $L_2$ spectrum on the magnitude of the experimental value of 
the spin-orbit splitting (0.6 eV for the Si $L_{2,3}$) and applying the intensity scaling based on the intensity ratios 2:4 
for L spectra.
Finally, to mimic the experimental XES we have considered an additional broadening of the theoretical spectra caused by
the apparatus function and the natural width of core levels. 
In the present work the natural width of the Si 2p was taken equal to 0.015 eV following 
the data published by Krause and Fuggle\cite{Krause79,Fuggle} and the apparatus function has been modeled  by the Gauss distribution with 
the full width half maximum of 0.3 eV. 
The details of calculation of such broadening are given in the Supporting Information.
It should be mentioned that in the implemented broadening procedure of the
calculated XES we have neglected the Auger broadening of the valence states. 
It is well-known that the broadening of valence states contribute  to the so-called "low energy tailing" of XES.

%It should be mentioned that in the implemented broadening procedure
%of the calculated XES we have neglected the broadening of the valence states.
%It is well-known that the broadening of valence states together with the "life-time broadening" 
%of the core states contribute both to the so-called "low energy tailing" of XES.

\section{Results and discussion}
\begin{figure}[htbp]
\centering
\includegraphics[scale=0.8]{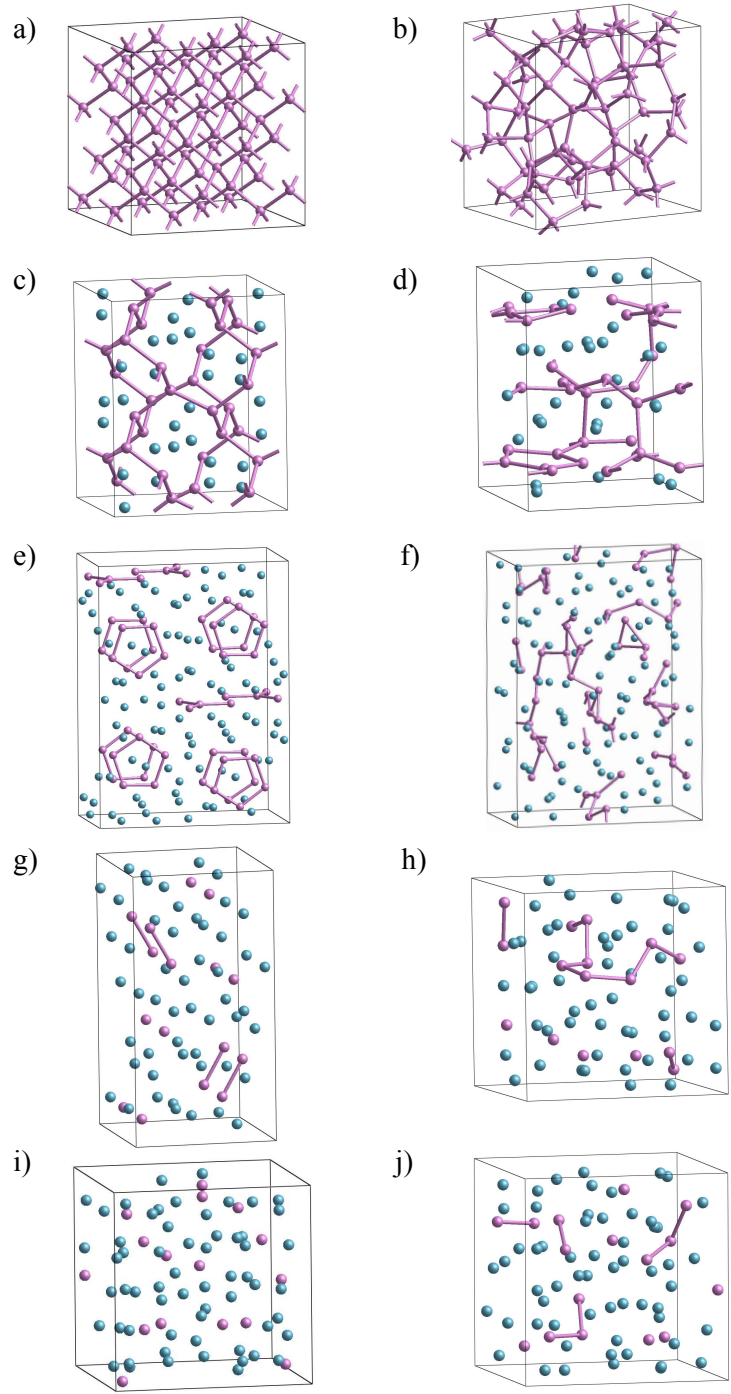}
\caption{Optimized structures of the crystalline and  amorphous phases of \ce{Si} (a) and (b), 
and \ce{Li_{x}Si} alloys: \ce{LiSi} (c) and (d), \ce{Li_{12}Si_{7}} (e) and (f), \ce{Li_{13}Si_{4}} (g) and (h),
\ce{Li_{15}Si_{4}} (i) and (j).
Silicon atoms are dull violet colored and lithium atoms are dull green/blue.}
%Four independent structures have been obtained for \ce{Li_{13}Si_{4}} and \ce{Li_{15}Si_{4}} for good statistics.}
\label{fig:structures}
\end{figure}

Optimized structures of the crystalline and amorphous phases of silicon as well as 
\ce{LiSi}, \ce{Li_{12}Si_{7}}, \ce{Li_{13}Si_{4}}, and \ce{Li_{15}Si_{4}} lithium silicides with the different concentration of Li atoms are presented in 
Figure \ref{fig:structures}. Crystalline silicon c-Si adopts the diamond cubic crystal structure, which  
presents an ordered network of tetrahedrally bonded Si atoms as shown in Figure \ref{fig:structures}a 
for the c-Si 2x2x2  supercell. The c-\ce{LiSi} structure possesses I41/a(88) space group and contains 
interconnected spiral chains of Si atoms in Li matrix, with the Si-Si bond length of 2.43 \AA\ within 
the chain and 2.52 \AA\ between the chains, as shown in Figure \ref{fig:structures}c for 1x1x2 supercell. 
Each of the Si atoms is coordinated with two neighboring atoms in the chain and one Si atom from the nearest chain. 
Therefore Si atoms in  c-\ce{LiSi} are three-fold coordinated. 
Coordination of Si atoms in \ce{Li_{x}Si} alloys decreases further with an increase in lithium concentration.
Thus c-\ce{Li_{12}Si_{7}} structure (Pnma(62) space group)  contains a combination of pentagonal Si rings, 
where Si atoms are two-fold coordinated 
and four atomic star-like Si clusters where the central Si atom is three-fold coordinated, as it is shown in Figure \ref{fig:structures}e. 
The average coordination of Si in  c-\ce{Li_{12}Si_{7}} is 1.86. 
The  c-\ce{Li_{13}Si_{4}} structure (Pbam(55) space group) contains Si atoms and dimers, with the average coordination 
number 0.5 (Figure \ref{fig:structures}g).
Finally, in the Li rich  
c-\ce{Li_{15}Si_{4}} I$\bar{\rm 4}$3d(220) structure Si atoms are scattered in the Li matrix at the large distances from each other
without formation of the covalent bonds (Figure \ref{fig:structures}i).

In amorphous structures the ordered Si-network disintegrates. In the a-\ce{Si} structure Si atoms are still 
coordinated with the 4 neighboring Si atoms, while in the case of the a-\ce{LiSi} structure the average Si coordination 
number (calculated with the cut off distance 2.6 \AA)  
decreases to 2.59. Thus silicon atoms in amorphous a-\ce{LiSi} are less coordinated than in the crystalline c-\ce{LiSi}.
However, in the case of the Li rich 
structures the Si coordination slightly increases if compared with the corresponding 
crystalline phases up to 2.07, 1.13 and 0.75 for a-\ce{Li_{12}Si_{7}}, a-\ce{Li_{13}Si_{4}}  and a-\ce{Li_{15}Si_{4}}, respectively.
Figures \ref{fig:structures}f,  \ref{fig:structures}h and \ref{fig:structures}j demonstrate that 
the a-\ce{Li_{12}Si_{7}} structure contains fragments of linear chains, triangles and tetramers,
the a-\ce{Li_{13}Si_{4}} structure contains silicon chains, atoms and dimers,
while in the a-\ce{Li_{15}Si_{4}} structure mostly the silicon atoms, dimers and trimers are observed.
As a general trend coordination of Si decreases with increase in Li concentration both for the crystalline and amorphous phases.
However, even at the high Li concentration  in amorphous  a-\ce{Li_{x}Si} alloys Si tends to form Si-Si covalent bonds in a good agreement 
with the results of previous theoretical calculations.\cite{Kim11,Chiang16}  
It is well known that soft X-ray  Si-L$_{2,3}$ emission is strongly 
affected by the local environment and hence it should be very sensitive to the coordination of Si atoms.
Therefore we suggest that such spectra carry information on the structural properties of the \ce{Li_{x}Si} alloys
and can be used for the comprehensive analysis of mechanisms of silicon lithiaton and disintegration of Si network 
into clusters of different sizes with increase in Li concentration.

\begin{figure}[htbp]
\centering
\includegraphics[scale=1]{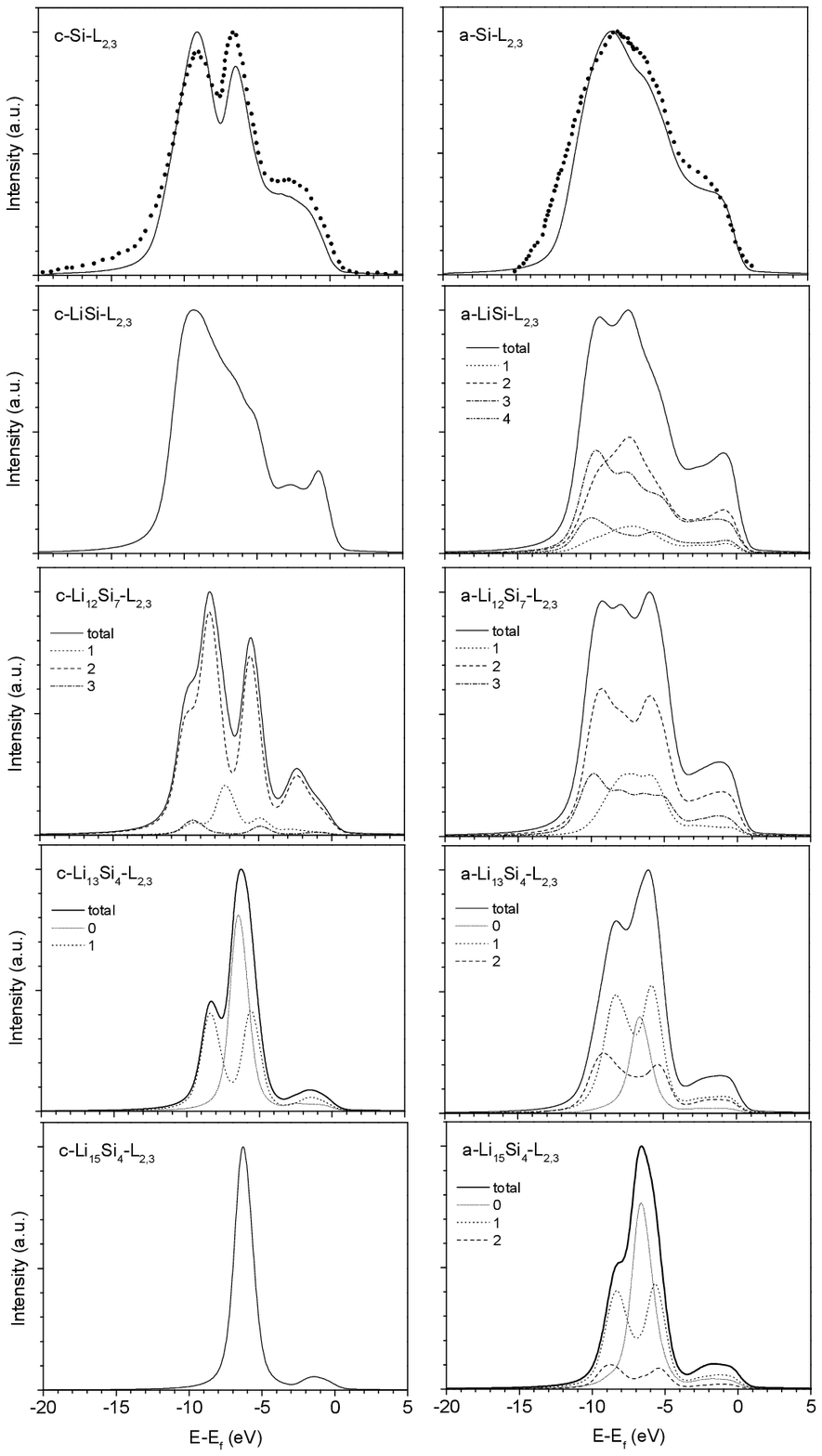}
\caption{Soft X-ray Si-L$_{2,3}$ emission spectra calculated for the crystalline 
and amorphous \ce{Si}, \ce{LiSi}, \ce{Li_{12}Si_{7}},  \ce{Li_{13}Si_{4}}, and \ce{Li_{15}Si_{4}} structures.
Numbers 0, 1, 2, and 3 denote emission from the isolated, single, double and triple coordinated Si atoms, respectively.
Black dots: experimental spectra of c-\ce{Si}\cite{Wiech73} and a-\ce{Si}\cite{Scimeca91} shifted by 98.65 eV 
to adjust the edge of the Si-L$_{2,3}$ emission band to the Fermi level.}
\label{fig:x-ray}
\end{figure}

\begin{figure}[htbp]
\centering
\includegraphics[scale=0.25]{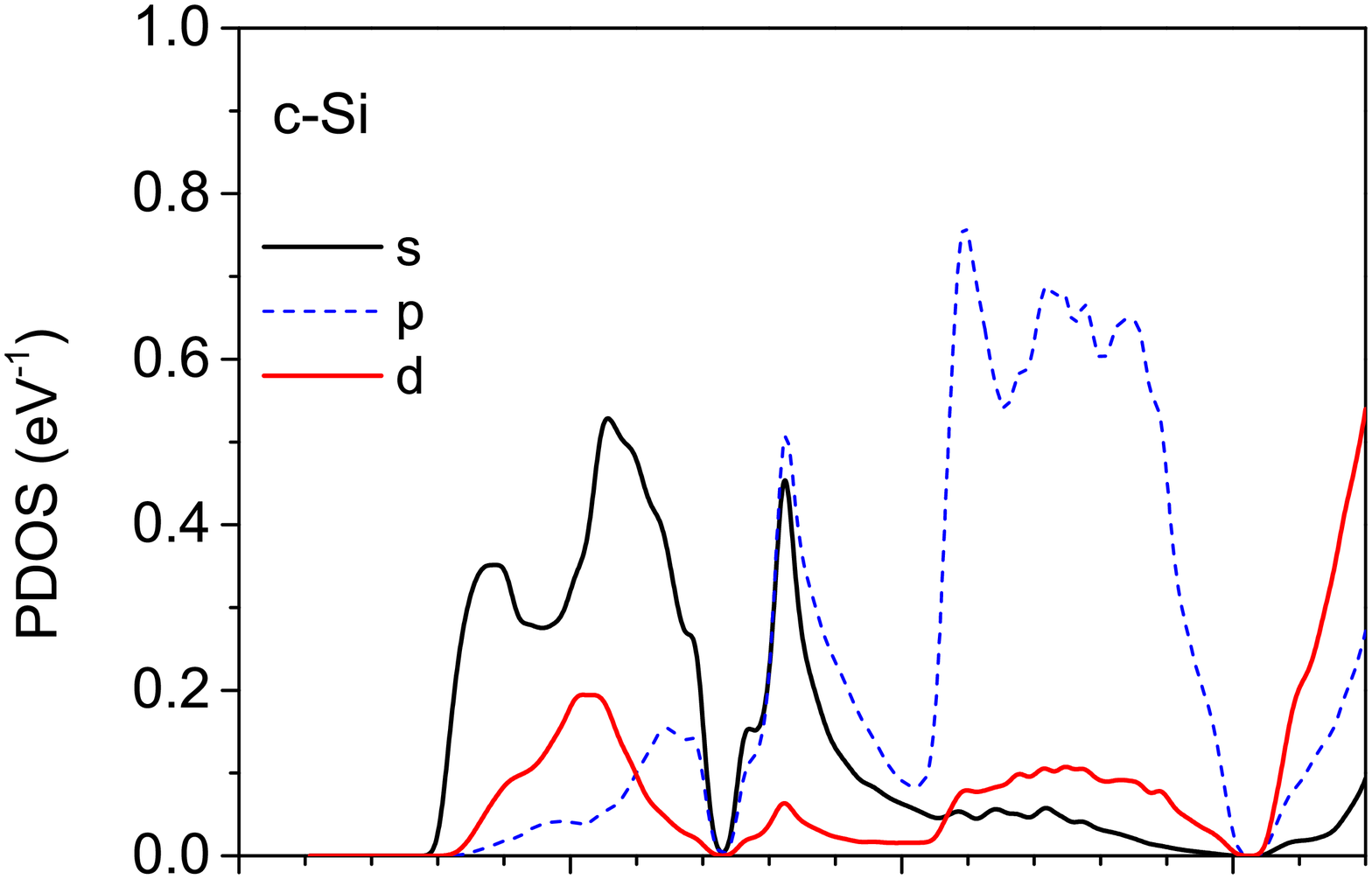}
\includegraphics[scale=0.25]{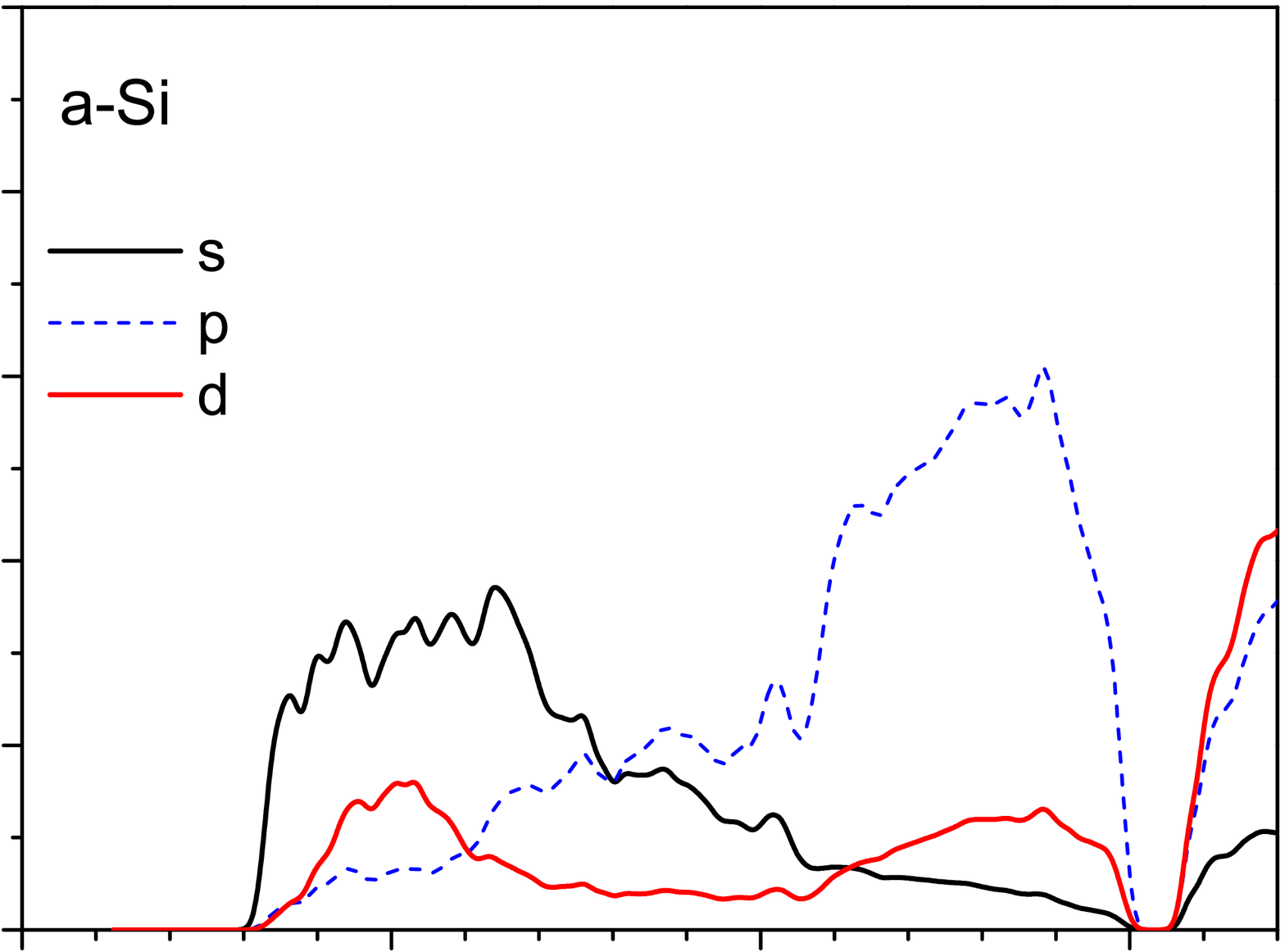}
\includegraphics[scale=0.25]{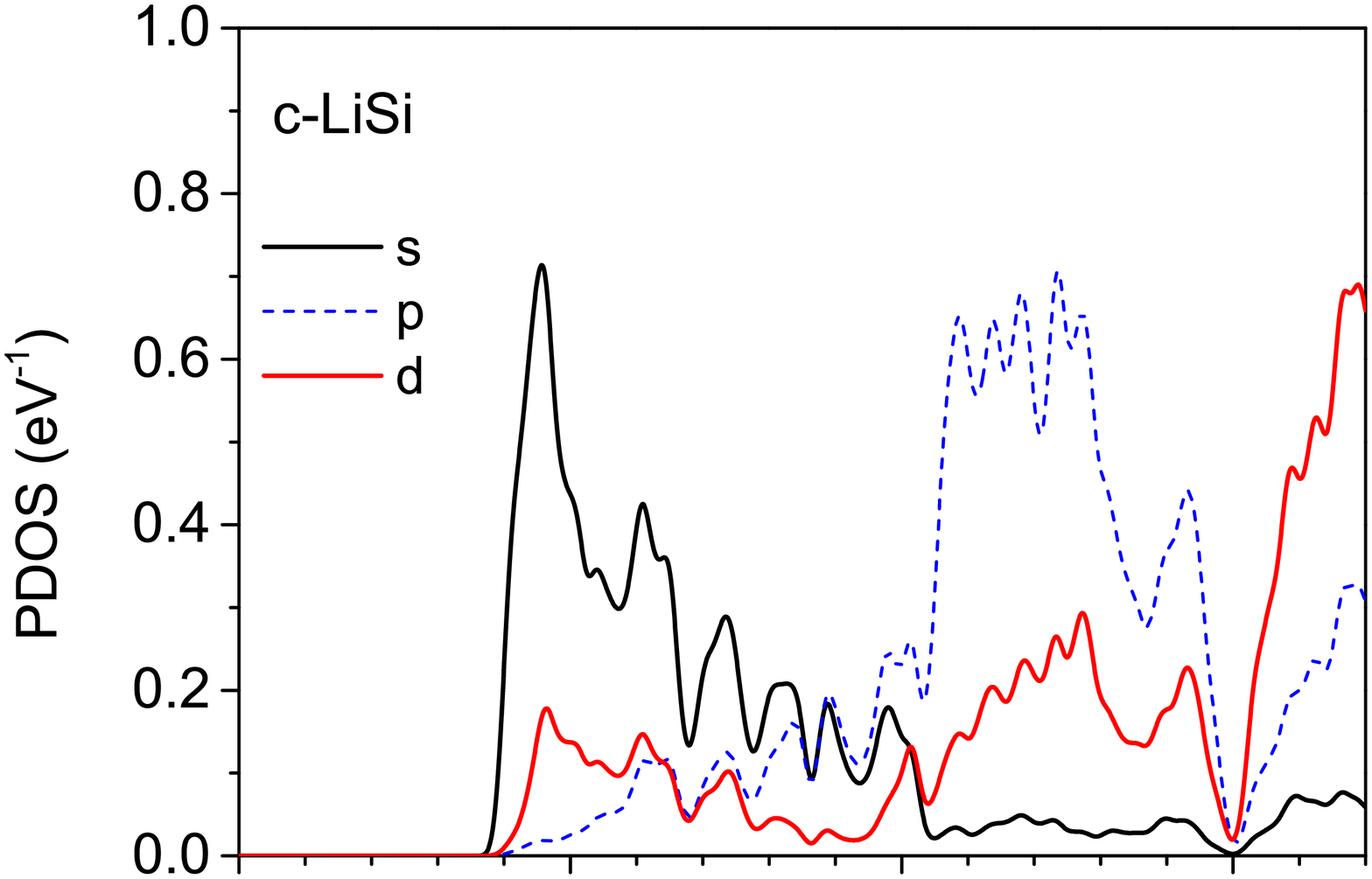}
\includegraphics[scale=0.25]{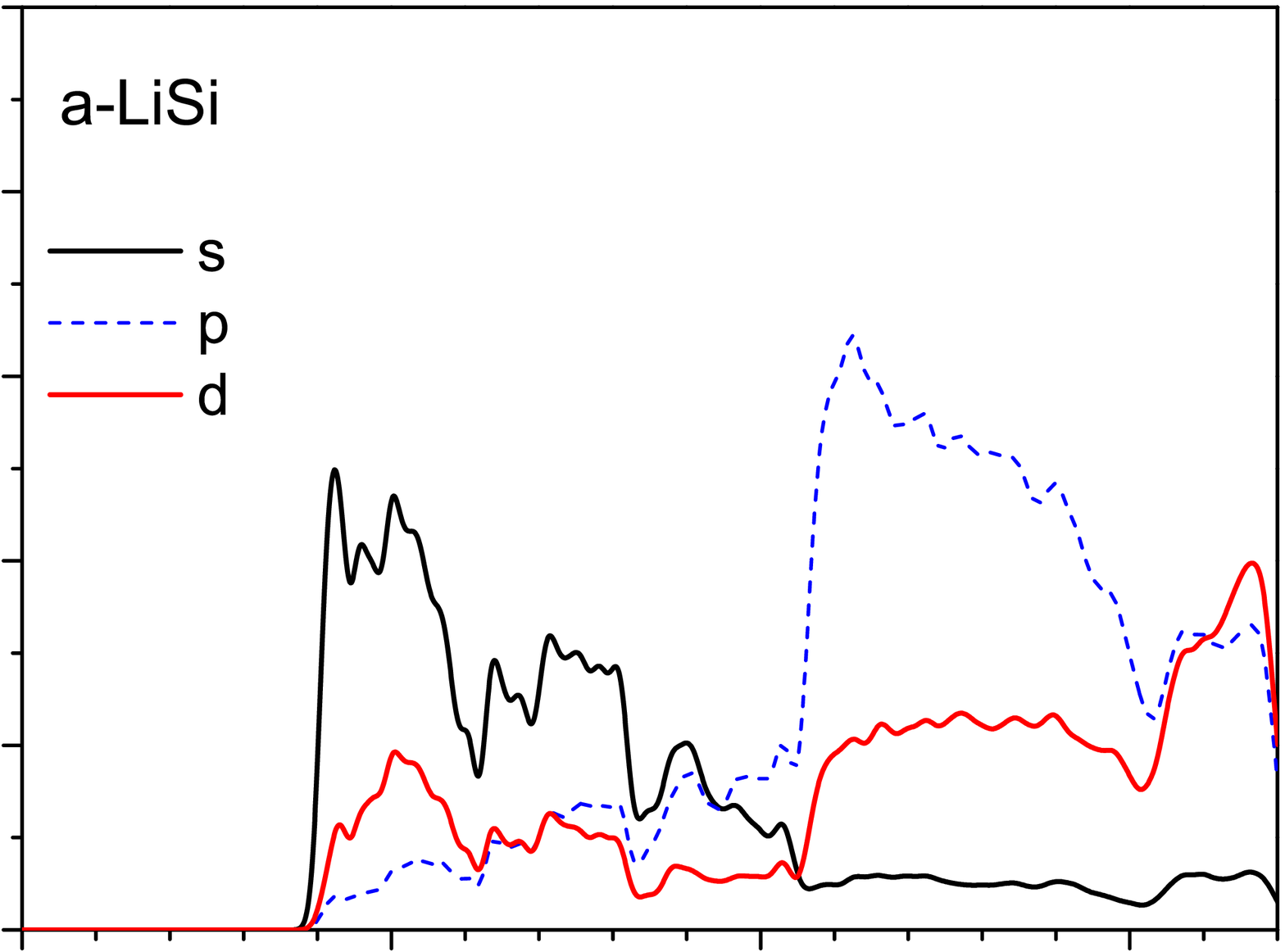}
\includegraphics[scale=0.25]{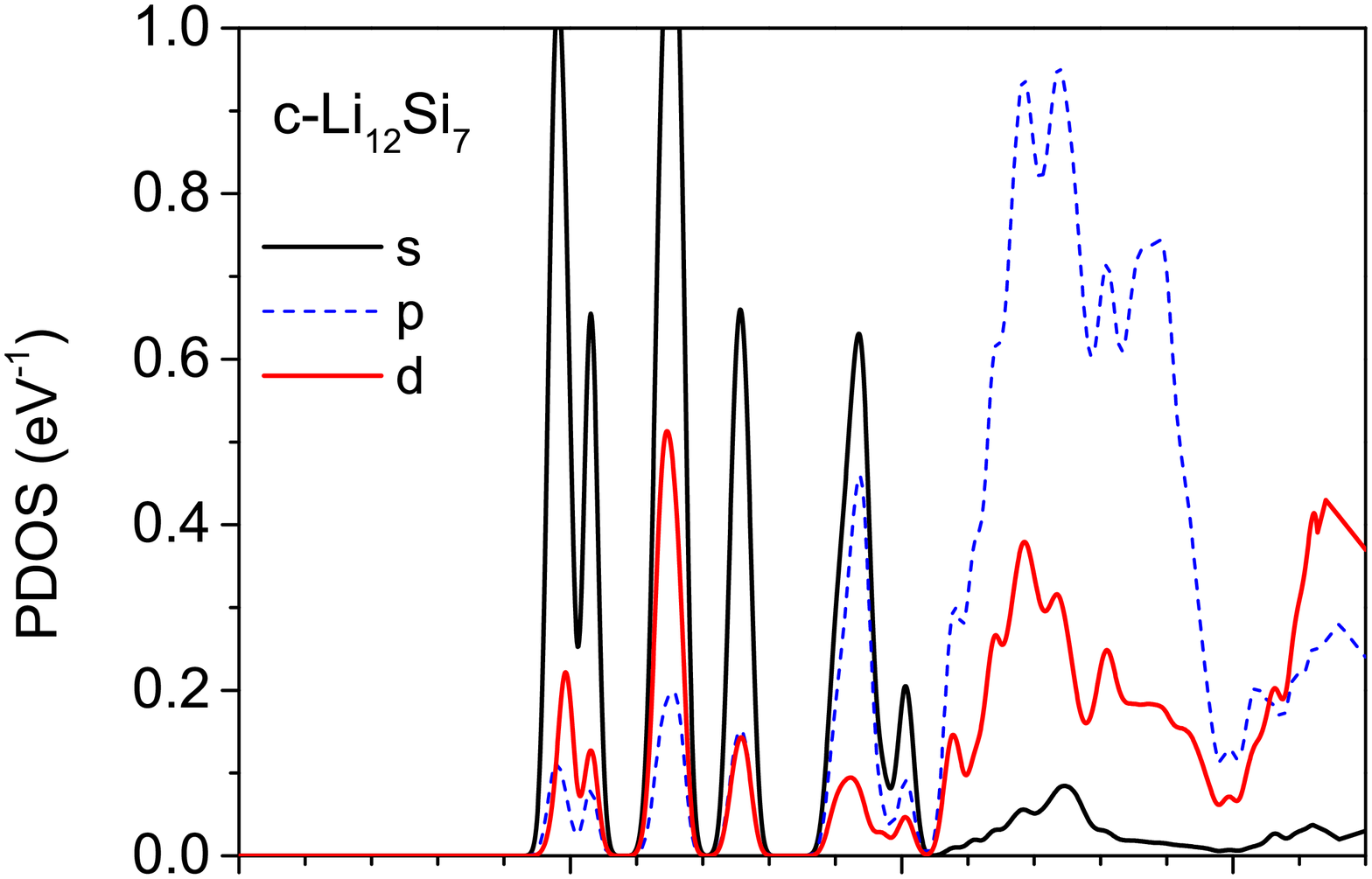}
\includegraphics[scale=0.25]{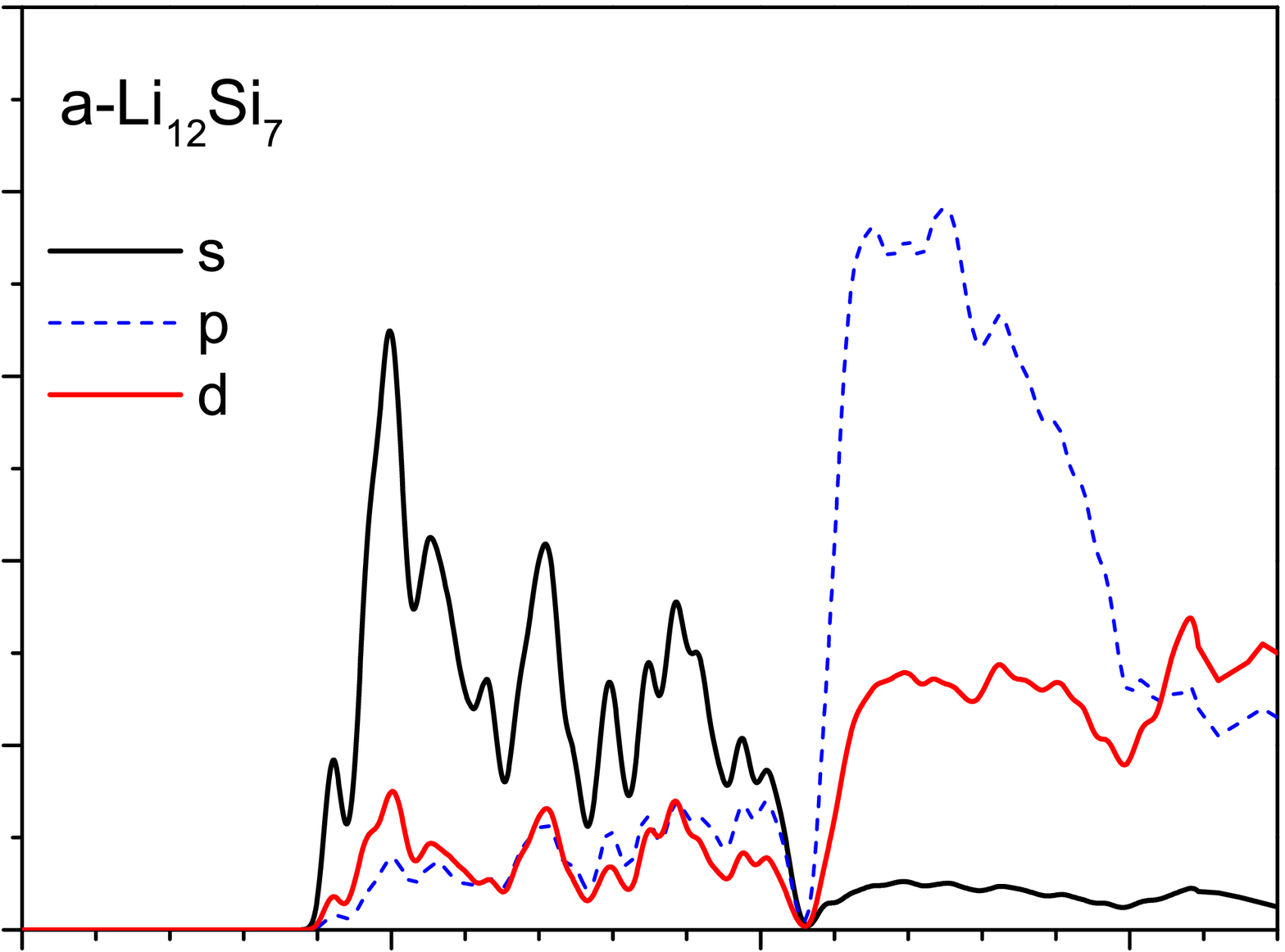}
\includegraphics[scale=0.25]{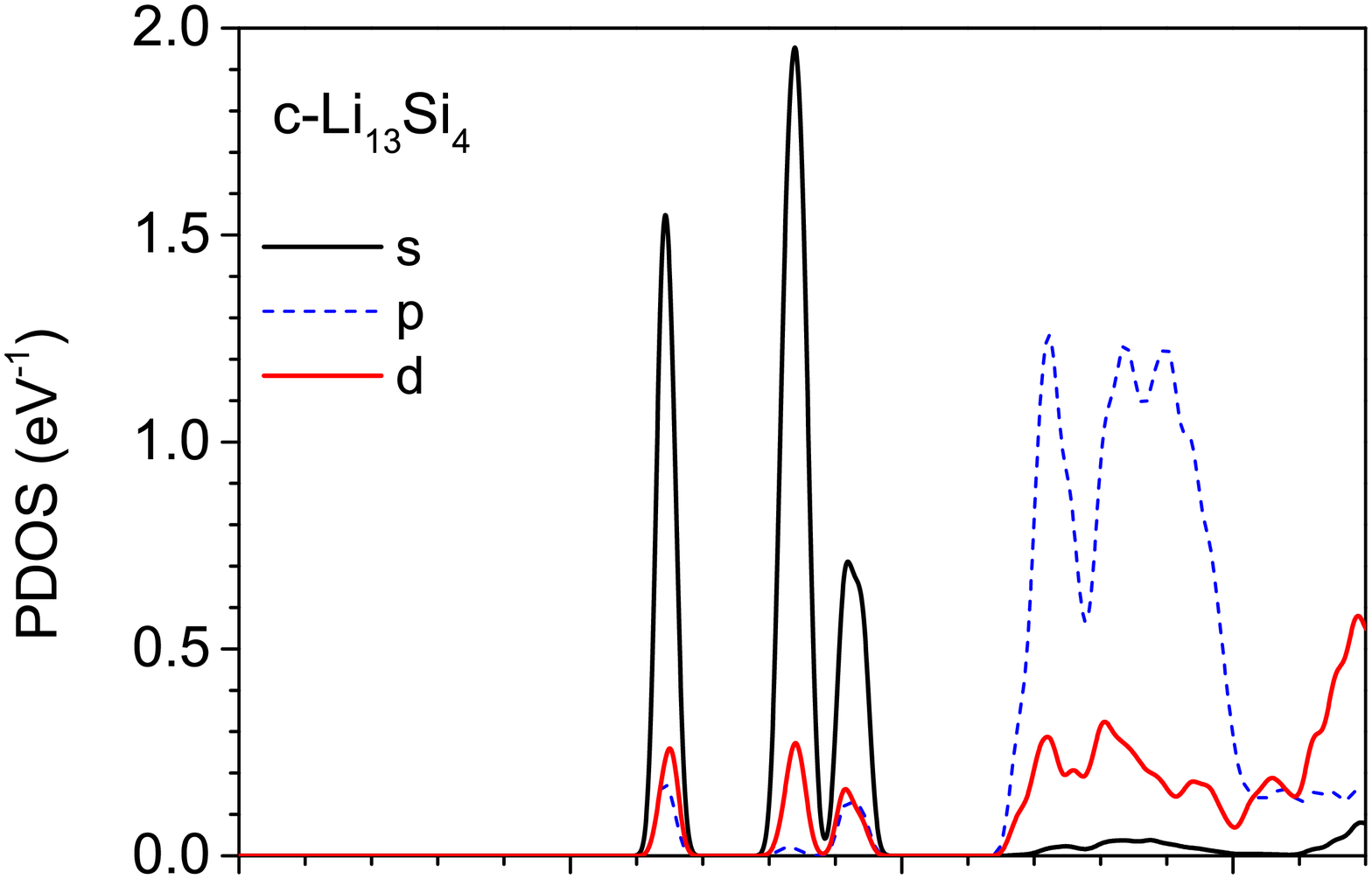} 
\includegraphics[scale=0.25]{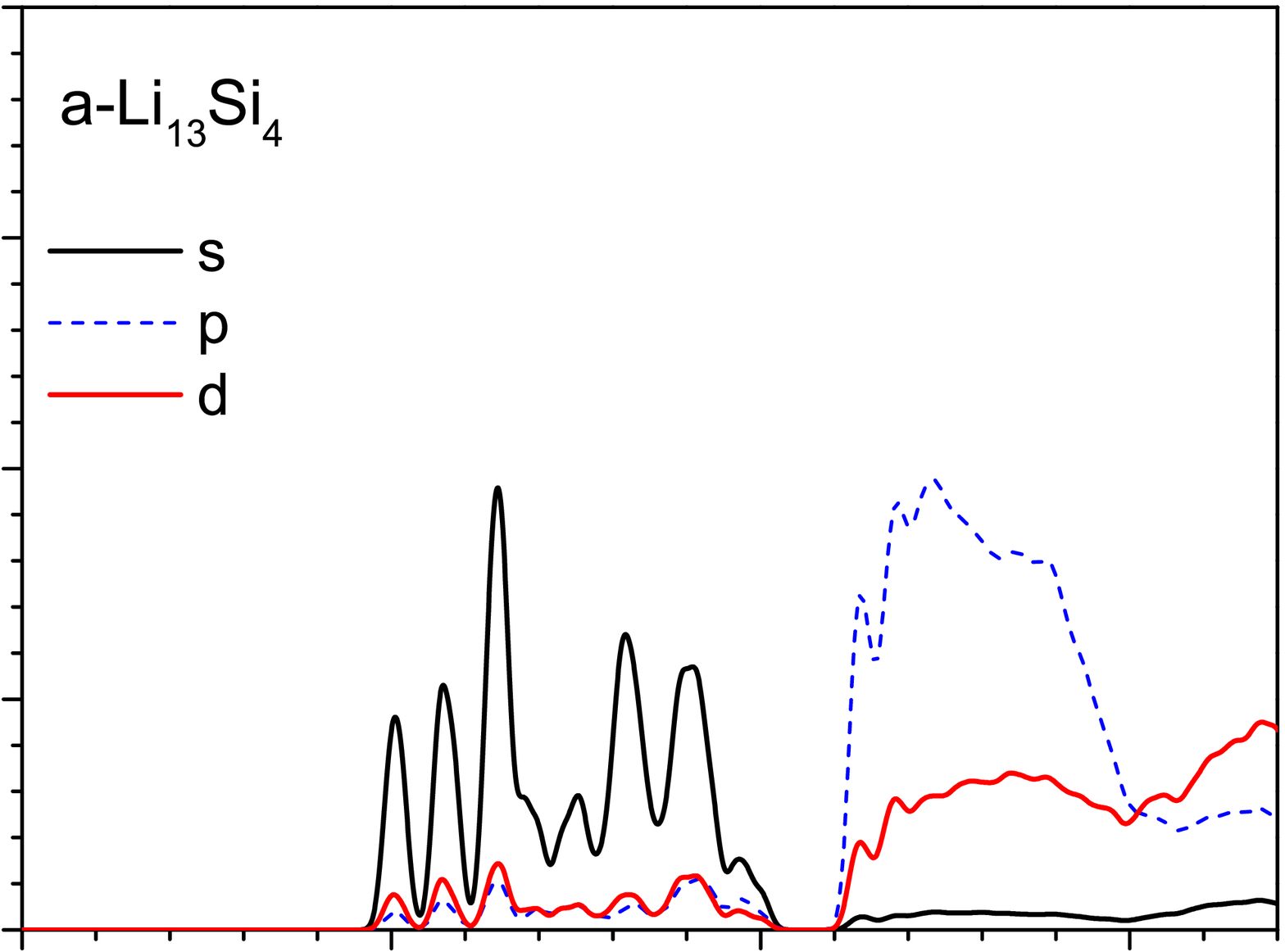}
\includegraphics[scale=0.25]{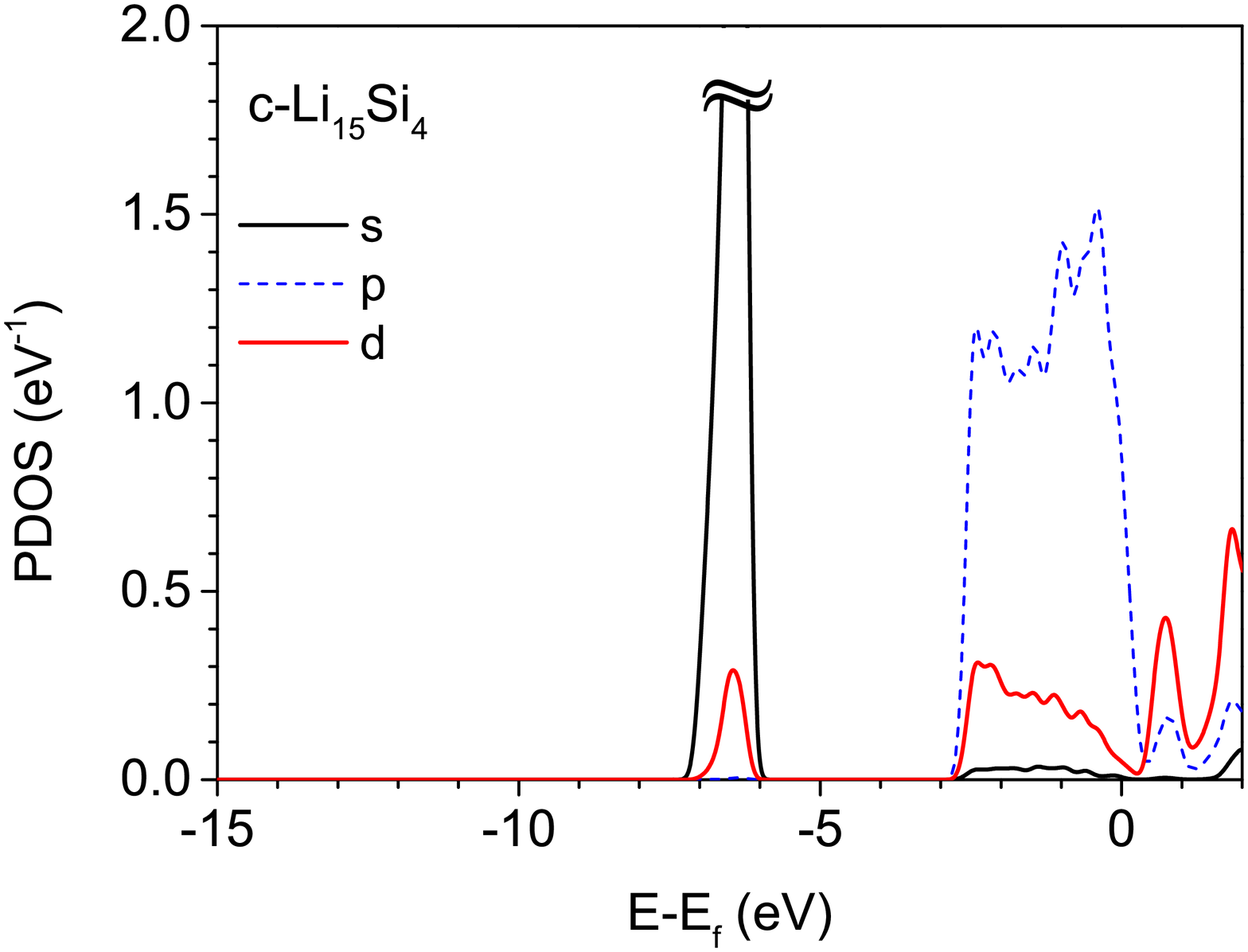}
\includegraphics[scale=0.25]{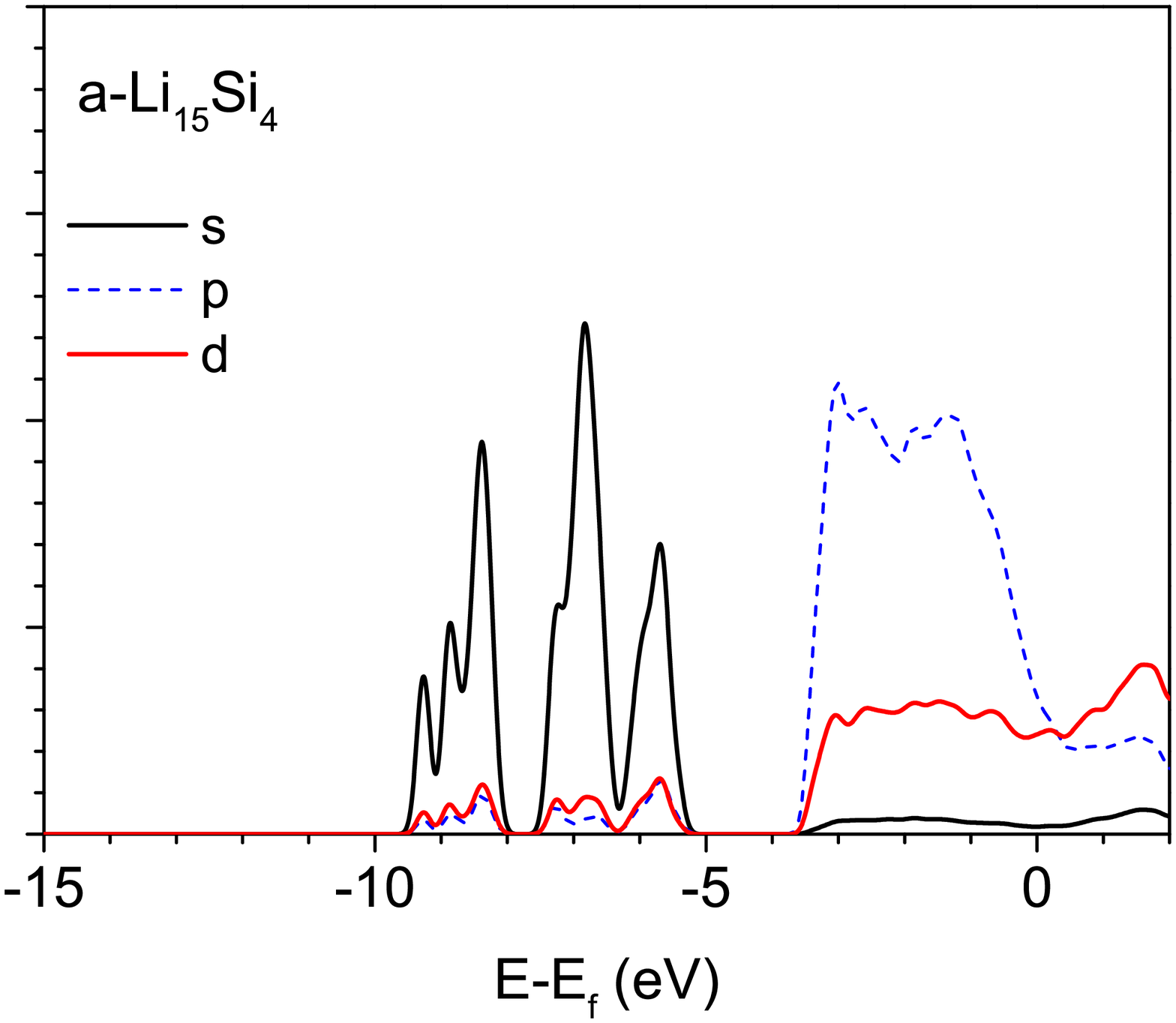}
\caption{Partial density of electronic states (PDOS) projected on Si atoms in crystalline 
and amorphous \ce{Si}, \ce{LiSi}, \ce{Li_{12}Si_{7}}, \ce{Li_{13}Si_{4}}, and \ce{Li_{15}Si_{4}} structures.
Solid black, dashed blue and solid red curves represent contribution of the s, p, and d states, respectively.
Gaussian broadening of half-width 0.1 eV has been used.}
\label{fig:DOS}
\end{figure}

Figure \ref{fig:x-ray} shows soft X-ray Si-L$_{2,3}$ emission spectra calculated for the crystalline and amorphous 
\ce{Si},  \ce{LiSi}, \ce{Li_{12}Si_{7}}, \ce{Li_{13}Si_{4}}, and \ce{Li_{15}Si_{4}} structures. 
The Si-L$_{2,3}$ emission spectra of the pure Si have been intensively studied both 
experimentally\cite{Ershov66,Wiech73,Scimeca91,Mashin01} and theoretically.\cite{Klima70}
Experimental L$_{2,3}$ spectrum of c-\ce{Si} exhibits two pronounced peaks at the photon energies 89.95 eV and 92.05 eV and broad 
shoulder at 95.2 eV with the fast decrease in slope at 97.5 eV.\cite{Wiech73} These photon energies correspond to the energies   
-8.70 eV, -6.60 eV, and -3.45 eV with respect to the 
the edge of the emission band which corresponds to the Fermi level. 
Figure \ref{fig:x-ray} demonstrates that the 
Si-L$_{2,3}$ emission spectrum of c-\ce{Si} calculated in this work reproduces very well 
the experimental one\cite{Wiech73} represented by dots. 
Note that in Figure \ref{fig:x-ray} the experimental spectrum is shifted by 98.65 eV to adjust the edge of the Si-L$_{2,3}$ emission 
band to the Fermi level. 

We would like to emphasize that the initial state of an X-ray radiative transition is the state of an atom with a vacancy in the 
core shell, which can be described by the atomic orbital $\phi_c({\bf r})$ of a free atom 
with the quantum numbers $n_c$, $l_c$ and $m_c$. As a result of the one-electron radiative
transition, a vacancy is formed in the valence band, which is described by the one-electron wave function $\psi_{n{\bf k}}({\bf r})$.
In the case of $l_c \geq 1$, the probability (\ref{eq:2}) of this transition can be separated into the partial contributions. 
Thus, in the case of the Si L$_{2,3}$ X-ray emission spectra (radiative electron transition to 2p vacancy of Si atom)   
the energy distribution of s- and d-states of the valence band mainly localized nearby the Si atom is reflected.
Therefore, in order to understand formation mechanisms of the soft X-ray Si-L$_{2,3}$ emission bands 
we have calculated the partial density of electronic states (PDOS) projected on Si atoms in the considered 
crystalline and amorphous \ce{Li_{x}Si} alloys as shown in Figure \ref{fig:DOS}. 
It is clearly seen that the low-energy peak in the XES of 
c-\ce{Si} is mainly associated with the 
low-lying valence 3s states in the energy range of  -12 eV -- -8 eV, as shown in  
Figure \ref{fig:DOS} by black line.
The second peak in the emission spectrum 
of c-\ce{Si} corresponds to the valence states with s-p hybridization, 
represented by a sharp peak in PDOS with the maximum at -6.78 eV. 
The high energy broad peak at -3.45 eV in the Si-L$_{2,3}$ X-ray emission from c-\ce{Si}  
is associated with the transitions from the 3d valence states 
appearing  at the energies of -4 -- -1 eV as a result of p-d hybridization.  

In the case of amorphous silicon the  Si-L$_{2,3}$ band exhibits a wide maximum at the photon energies 90.5 eV (-8.15 eV) with a shoulder at 
$\sim$ 96.5 eV (-2.15 eV)\cite{Scimeca91} as it is shown in Figure \ref{fig:x-ray}. Here numbers in parentheses correspond 
to the energy scale shifted by 98.65 eV to adjust the edge of the Si-L$_{2,3}$ emission band to the Fermi level. 
The main Si-L$_{2,3}$ emission band of the a-\ce{Si} becomes broader if compared with the one for c-\ce{Si} and the second peak
in the spectrum corresponding to the s-p hybridized states disappears. 
Indeed, PDOS of a-\ce{Si} presented in Figure \ref{fig:DOS} demonstrates  absence of the s-p hybridization. 
Therefore in the Si-L$_{2,3}$ spectrum of a-\ce{Si} the main peak corresponds to the transitions from the s states of the valence band,
while the high energy shoulder maps the d states of the valence band. The  calculated  Si-L$_{2,3}$ spectrum 
of a-\ce{Si} excellently reproduces the experimental one,
showing that the theoretical approach used in the present work is very reliable. 

Lithiation considerably affects the Si-L$_{2,3}$ emission spectra  of the crystalline and amorphous silicon.
The emission band of the c-\ce{LiSi} exhibits the low energy asymmetric peak with the maximum at -9.35 eV below the Fermi level, 
as shown in Figure \ref{fig:x-ray}. The right shoulder of the peak has complicated structure and drops sharply at -4 eV. 
The energy position of this peak is very close to the low energy peak in the  Si-L$_{2,3}$ spectra of c-Si. 
It is seen from the PDOS of c-\ce{LiSi} (Figure \ref{fig:DOS}) that the low energy Si-L$_{2,3}$ emission band is mainly formed 
by the radiation transitions from the s valence states where a sharp low energy peak is followed 
by the five peaks gradually decreasing in intensity as energy increases up to -5 eV.  
The high energy part of the emission spectra of c-\ce{LiSi} consists of 
a wide maximum at -2.7 eV and more sharp maximum at -0.8 eV. These two peaks originate from the d states of 
the valence band of c-\ce{LiSi} lying in the energy range of -4 eV -- 0 eV as it is seen from Figure \ref{fig:DOS}.
The change in the shape of the Si-L$_{2,3}$ emission band of c-\ce{LiSi} in comparison with the pure c-\ce{Si} is related  
to the change of the local environment of the Si atoms due to lithiation, which affects the density of the valence states of c-\ce{LiSi}. 
As it was mentioned above Si atoms in c-\ce{LiSi} are coordinated with three nearest Si atoms, 
while the coordination number in c-\ce{Si} crystal is four.
One can think that amorphization of \ce{LiSi} structure would result in the smoothening of the Si-L$_{2,3}$ emission band shape.
However, Figure \ref{fig:x-ray} demonstrates that the low energy Si-L$_{2,3}$ band of a-\ce{LiSi} 
has pronounced double peak structure with maxima at -9.4 eV and -7.4 eV. The similar structure also appears in the PDOS of 
the s states of the valence band of a-\ce{LiSi}. The pronounced high energy peak in the emission spectrum at -0.9 eV is formed 
by the radiation transitions from the d states of the valence band.  
In order to understand the mechanism of formation of the double peak 
structure in the low energy  Si-L$_{2,3}$ band of a-\ce{LiSi} we decomposed the total emission spectrum over 
contributions from the Si atoms with different coordination. In Figure \ref{fig:x-ray} the contributions 
from the groups of single-, two-, three-, and four-fold coordinated Si atoms are presented by 
dotted, dashed, dash-doted and dash-double-dotted lines, respectively. As it is seen from Figure \ref{fig:x-ray}
the two- and three-fold  coordinated Si atoms give the main contribution to the total spectrum. 
The spectral distribution of the emission from the two-fold coordinated Si atoms in a-\ce{LiSi}
has a maximum at -7.4 eV, while the main peak in the emission from the three-fold coordinated atoms 
is located at -9.5 eV. Therefore the double peak structure in the main Si-L$_{2,3}$ emission band of a-\ce{LiSi} can be 
explained by the superposition of the photon emission from the silicon atoms with the different coordination.  

The Si-L$_{2,3}$ emission spectrum of c-\ce{Li_{12}Si_{7}} consists of  the three well separated maxima at -8.4 eV, -5.6 eV and -2.4 eV.
The high energy peak is formed by the transition from the d states of the valence band, while the two low energy 
peaks are formed mostly by the s states, presented in PDOS by a group of the energetically well separated sharp lines 
at the energies -10 eV -- -5 eV.  The two-fold coordinated Si atoms give dominant contribution to the whole spectrum.
In the case of the amorphous a-\ce{Li_{12}Si_{7}} lithium silicide the Si-L$_{2,3}$ emission spectrum 
shows more complicated structure with 
three peaks in the low energy s band at energies -9.3 eV, -8.0 eV and -6.0 eV 
and a broad maximum at -1.2 eV corresponding to the d band. The peaks at -9.3 eV and -6.0 eV  are mostly formed 
by the emission from the double coordinated Si atoms with some small contribution from the triple coordinated Si atoms, 
while the peak in the middle of the band at -8.0 eV is formed  by the emission from the single coordinated atoms, 
as shown in Figure \ref{fig:x-ray}. 

As it was discussed above the c-\ce{Li_{13}Si_{4}} structure contains an equal number of the isolated Si atoms and 
Si atoms forming dimers. The X-ray emission from these two groups of atoms shows different spectral dependence. 
The Si-L$_{2,3}$ emission spectrum from the isolated atoms possesses sharp line, while the spectrum from dimers demonstrates the
double peak structure, as shown in Figure \ref{fig:x-ray}. Therefore 
the total emission spectrum of c-\ce{Li_{13}Si_{4}} shows sharp peak at -6.3 eV which corresponds 
mainly to the emission from the isolated Si atoms and the low energy peak at -8.3 eV  corresponding to the 
radiation from the single coordinated Si. In the amorphous a-\ce{Li_{13}Si_{4}} structure x-ray emission
from the single and double coordinated Si atoms dominates over the one from the isolated Si. Therefore 
the main Si-L$_{2,3}$ emission band of a-\ce{Li_{13}Si_{4}} becomes broader in comparison with the spectrum 
of the corresponding crystalline phase. The main peak in the Si-L$_{2,3}$ spectrum a-\ce{Li_{13}Si_{4}} at -6.0 eV corresponds 
to the emission from the single coordinated atoms.
  
In the case of c-\ce{Li_{15}Si_{4}} structure all Si atoms are separated without formation of the Si--Si covalent bonds.
Therefore the Si-L$_{2,3}$ emission spectrum of c-\ce{Li_{15}Si_{4}} possesses only one sharp intensive spectral line at -6.3 eV  
which corresponds to the transitions from the s states of Si and the low intensity small bump at -1.4 eV reflecting 
small presence of the d states in the valence band. On the other hand the emission spectra of the amorphous  a-\ce{Li_{15}Si_{4}}
demonstrate low energy structure with pronounced peaks at -8.3 eV and -6.5 eV corresponding 
to radiation from the single coordinated and isolated Si atoms, respectively. Small contribution from the double coordinated 
Si atoms with a double peak structure at -8.8 eV and -5.4 eV is also noticeable.  Presence of the 
singly and double coordinated atoms in the amorphous a-\ce{Li_{15}Si_{4}} structure results in broadening of the 
the emission band. The high energy wide peak at -1.3 eV corresponds to the transitions from the d states localized in the vicinity of the 
single coordinated Si atoms. 

Our calculations demonstrate that the Si-L$_{2,3}$ emission spectra of the crystalline and amorphous 
\ce{Li_{x}Si} alloys studied in this work show very different spectral dependencies reflecting the
difference in Si-Si interaction in these systems as well as the difference in disintegration of the Si-Si network into Si clusters 
and chains of the different sizes upon silicon lithiation. The low energy Si-L$_{2,3}$ emission bands of the \ce{Li_{x}Si} alloys
become narrower and shift towards higher energies (closer to the Fermi level) with increase in 
Li concentration. Therefore the soft X-ray emission  spectra of the \ce{Li_{x}Si} alloys 
can be used as a powerful tool for investigation of the silicon lithiation process and transition from the crystalline 
to amorphous structures in LIBs.

Recently, it was demonstrated experimentally that several well separated  
\ce{Li_{x}Si} phases are formed on Si(111) substrate upon electrochemical lithiation.\cite{Aoki16} 
The above experiments have been performed with the use of methods of the soft X-ray emission spectroscopy 
combined with the scanning electron microscopy and the X-ray diffraction.  
It was suggested that the first layer formed upon electrochemical Si(111) lithiation is  a single-crystalline 
c-\ce{Li_{15}Si_{4}} alloy phase, while the second layer is a mixture of amorphous 
a-\ce{Li_{15}Si_{4}} and/or a-\ce{Li_{13}Si_{4}} phases.\cite{Aoki16}
However the detailed interpretation of the observed results was difficult due to the absence 
of theoretical data. In the present work we fill this gap and perform theoretical analysis 
of the Si-L$_{2,3}$ emission spectra of the different phases of \ce{Li_{x}Si} alloys  by
comparing our results with the experimental XES reported by Aoki et al.\cite{Aoki16}

\begin{figure}[htbp]
\centering
\includegraphics[scale=0.35]{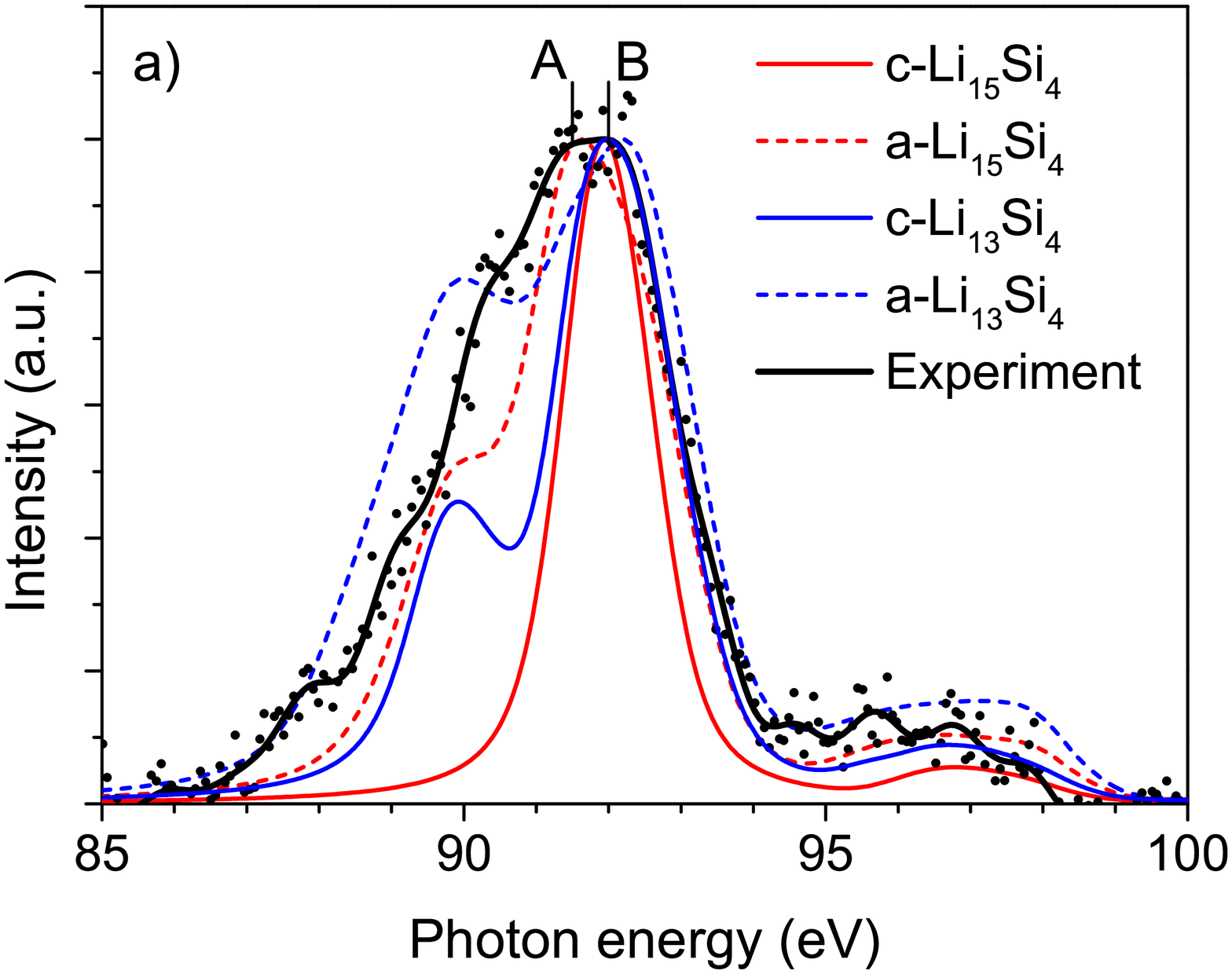}
\includegraphics[scale=0.35]{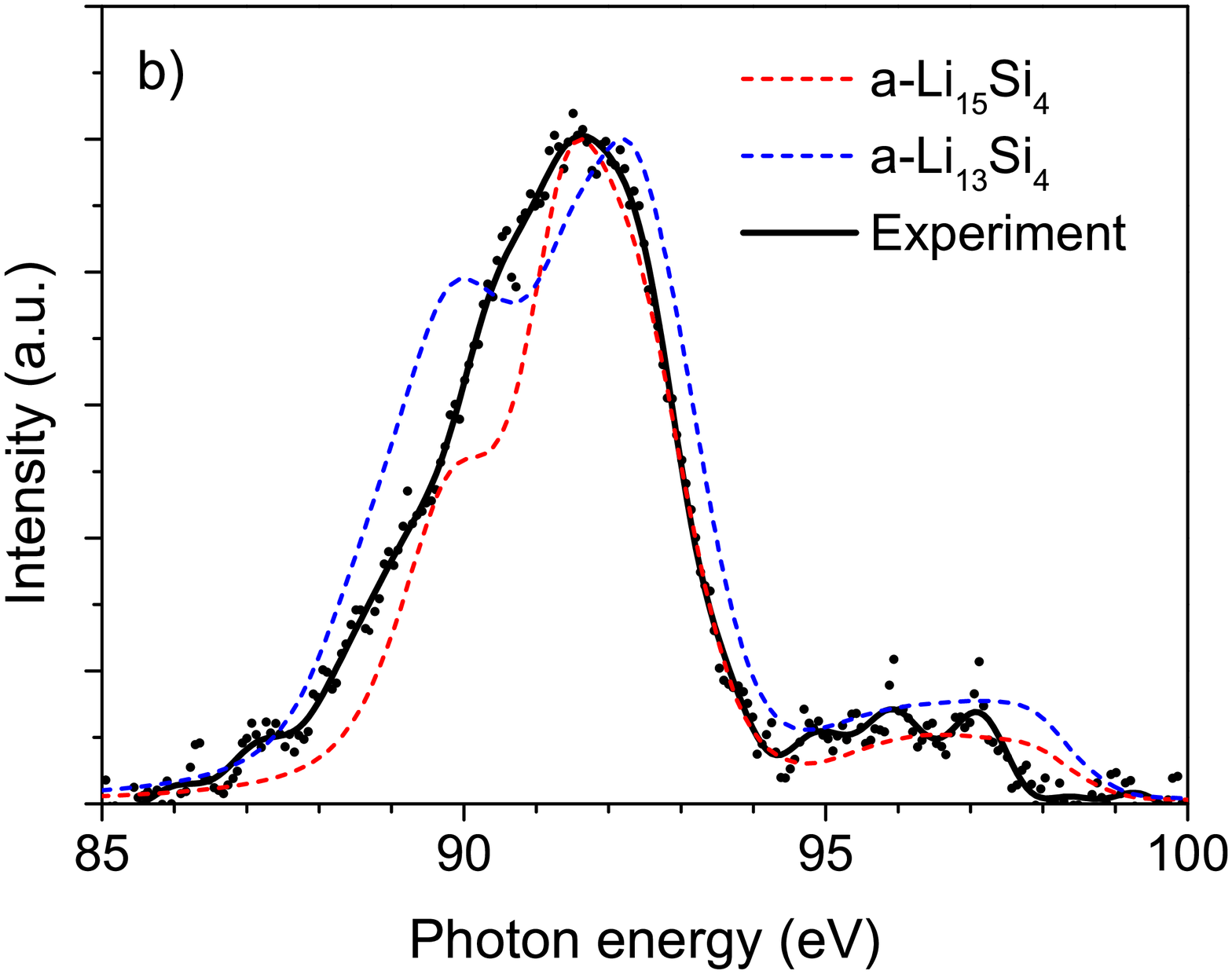}
\caption{FFT filtered experimental Si-L$_{2,3}$ emission specta of the first (a) and second (b) outermost layers of the 
electrochemically lithiated Si(111) - black lines; unfiltered experimental data - black dots.\cite{Aoki16}  
Theoretical Si-L$_{2,3}$ emission spectra of the  
c-\ce{Li_{15}Si_{4}} (red line), 
a-\ce{Li_{15}Si_{4}} (red dashed line) 
c-\ce{Li_{13}Si_{4}} (blue line), 
a-\ce{Li_{13}Si_{4}} (blue dashed line) 
shifted by  the experimental value\cite{Aoki16} (98.24 eV) of the edge energy of the Si-L$_{2,3}$ spectra 
of the first and second layers.}
\label{fig:experiment}
\end{figure}

The experimental Si-L$_{2,3}$ XES reported by Aoki et al.\cite{Aoki16}
for the first and second  outermost layers of the 
electrochemically lithiated Si(111)
are represented by black dots in Figures \ref{fig:experiment}a and  \ref{fig:experiment}b, 
respectively. 
In order to reduce an artificial noise from the spectra we perform
fast Fourier transform (FFT) filtering of the experimental data. 
The FFT filtered spectra are shown in Figures \ref{fig:experiment}a and  \ref{fig:experiment}b
by black lines.

It was suggested by  Aoki et al.\cite{Aoki16} that the experimental XES of the second layer of the electrochemically 
lithiated Si(111) corresponds to the amorphous a-\ce{Li_{15}Si_{4}} and/or a-\ce{Li_{13}Si_{4}} phases. 
Figure \ref{fig:experiment}b demonstrates that
the experimental XES of the second layer 
is in excellent agreement  with 
the theoretical spectrum of the a-\ce{Li_{15}Si_{4}} phase.
Note that theoretical spectra in Figure \ref{fig:experiment} are shifted by 98.24 eV 
which corresponds to the edge energy of Si-L$_{2,3}$ spectra of the first and second layers reported 
experimentally.\cite{Aoki16} 
One should note, that contribution of the a-\ce{Li_{13}Si_{4}} phase to the XES of the second layer can not be excluded completely, but  
lowering the concentration of lithium results in formation of the Si clusters of the larger sizes in the lithium matrix and 
hence broadening the Si-L$_{2,3}$ emission band.  Figure \ref{fig:experiment}b  demonstrates that the calculated 
XES of a-\ce{Li_{13}Si_{4}} is broader than experimental spectrum. Moreover, the energy position of the emission peak 
of a-\ce{Li_{13}Si_{4}} is 0.55 eV higher in energy if compared with the one of a-\ce{Li_{15}Si_{4}}.
Therefore it is unlikely that a-\ce{Li_{13}Si_{4}} gives noticable contribution to the XES of the 
second layer observed by  Aoki et al.\cite{Aoki16}

The analysis of the Si-L$_{2,3}$ emission of the first layer is more complicated. 
The scanning electron microscopy image of the first layer of the electrochemically lithiated Si(111) demonstrates 
presence of the regular triangular pyramids with an average height of 1.0 $\mu$m on the surface.\cite{Aoki16} 
This observation indicates manifestation of the crystalline structure.
Based on the analysis of the XRD patterns it was concluded that the first layer most likely consist of    
the crystalline c-\ce{Li_{15}Si_{4}} phase.\cite{Aoki16} 
However, comparison of the experimental XES of the first layer reported by Aoki et al.\cite{Aoki16} with the results of our 
theoretical calculations clearly demonstrates that the X-ray emission from the first layer can not be 
explained by the emission from the crystalline 
c-\ce{Li_{15}Si_{4}} phase  only because of the considerable difference 
in the width and shape of the main Si-L$_{2,3}$ band. As it was discussed above the c-\ce{Li_{15}Si_{4}} 
structure contains only isolated Si atoms, resulting
in formation of the narrow spectral line in Si-L$_{2,3}$ emission  as
shown in Figure \ref{fig:experiment}a by solid red line.

One can notice that the shape of the experimental XES
of the first layer in the vicinity of its maximum is flat,
reflecting superposition of two peaks at $\sim$91.5 eV and $\sim$92.0 eV marked 
in Figure \ref{fig:experiment}a as A and B, respectively.
The position of the peak B ideally fits to the energy position of the spectral line of c-\ce{Li_{15}Si_{4}} phase,
while position of the peak A excellently agrees with the maximum in the XES of 
a-\ce{Li_{15}Si_{4}} phase. Therefore we can conclude that XES of the first layer observed   
by Aoki et al.\cite{Aoki16} presents superposition of the emission of the 
crystalline and amorphous \ce{Li_{15}Si_{4}} phases. 
We should also note that energy position of the peak B fits 
the position of the theoretical spectra of the  c-\ce{Li_{13}Si_{4}} structure, 
as shown in Figure \ref{fig:experiment}a, therefore we can not exclude completely 
contribution from the c-\ce{Li_{13}Si_{4}} phase.
Note that the shoulder on the left slope of the main band of the experimental
spectrum excellently fits to the position of the low energy peak which corresponds 
to the emission from the single coordinated Si atoms in Si dimers or small Si chains.  
Such single coordinated Si atoms are absent in the c-\ce{Li_{15}Si_{4}} phase, but manifest themselves in the 
a-\ce{Li_{15}Si_{4}}, c-\ce{Li_{13}Si_{4}} and a-\ce{Li_{13}Si_{4}} phases.
Our calculations clearly demonstrate that Si-L$_{2,3}$ emission bands of 
the crystalline and amorphous  \ce{Li_{15}Si_{4}} and \ce{Li_{13}Si_{4}} 
structures are considerably different in terms of the shape and energy position 
and can be used for determination of the chemical structure and composition of 
the first layer of the electrochemically lithiated silicon. Therefore we believe 
that our work will stimulate intensive experimental investigation of the soft 
X-ray emission of  the \ce{Li_{x}Si} alloys in order to resolve puzzle of the first layer.

%Although we can not exclude completely presence of the emission from the a-\ce{Li_{13}Si_{4}} phase in the first layer, 
%it is quite unlikely, because  Si-L$_{2,3}$  band of a-\ce{Li_{13}Si_{4}} is wider and shifted 0.3 eV higher in energy
%if compared with the one from c-\ce{Li_{13}Si_{4}}. 

The analysis performed above clearly demonstrates that combination 
of the theoretical and experimental methods of the soft X-ray  emission spectroscopy can be used as a powerful tool 
for the comprehensive analysis of the anode materials in LIBs.

\section{Conclusion}
In conclusion, we have shown that the methods of the soft X-ray emission spectroscopy 
can be used as a powerful tool for the comprehensive analysis of the electronic and structural properties of 
the crystalline and amorphous \ce{Li_{x}Si} alloys forming in LIBs upon Si lithiation. 
DFT calculations of the crystalline and amorphous structures of \ce{Si} and  
\ce{LiSi}, \ce{Li_{12}Si_{7}}, \ce{Li_{13}Si_{4}}, and \ce{Li_{15}Si_{4}} alloys with the different concentration of Li atoms  
show that the Si coordination decreases with increase in Li concentration both for the crystalline and amorphous phases,
however even at the high Li concentration in amorphous  a-\ce{Li_{x}Si} alloys Si tends to cluster 
forming covalent Si-Si bonds.  
We have demonstrated  that Si-L$_{2,3}$ emission spectra of the crystalline and amorphous 
\ce{Li_{x}Si} alloys possess different spectral dependencies reflecting the
difference in Si-Si interaction in these systems as well as the difference in disintegration 
of the Si-Si network into Si clusters and chains of the different sizes.
Theoretical calculations predict that the Si-L$_{2,3}$ emission bands of the \ce{Li_{x}Si} alloys
become narrower and shift towards higher energies with increase in 
Li concentration. 
Comparison of the theoretical  Si-L$_{2,3}$ emission spectra with  the experimentally obtained XES from
the first two layers of the electrochemically lithiated Si(111) demonstrates that 
XES of the top layer can be explained by the superposition of the x-ray emission from the 
c-\ce{Li_{15}Si_{4}} and a-\ce{Li_{15}Si_{4}} phases, while the second layer mostly 
consists of the a-\ce{Li_{15}Si_{4}} phase.   
The shape and position of the Si-L$_{2,3}$ emission bands of \ce{Li_{x}Si} reflect
the relative contribution of the X-ray radiation from the 
Si atoms with different coordination which can be used for the detailed analysis of the 
lithiation process of Si in LIBs.

\begin{acknowledgement}
This work was supported by the Ministry of Education, Culture, Sports, Science and Technology (MEXT) of Japan for the  
program on the Development of Environmental Technology using Nanotechnology; 
AL, AN and TT gratefully acknowledge support from the Japan Society for the Promotion of Science 
(JSPS KAKENHI Grants 15K05387, 15K06563 and 16KT0047, respectively)
and partial support by MEXT as "Priority Issue on Post-K computer" 
(Development of new fundamental technologies for high-efficiency energy creation, conversion/storage and use);
IVA and VGK gratefully acknowledge support from the Russian Foundation for Basic Research (Grants  15-03-07543, 16-08-01244 and 18-03-00750).
IIT acknowledges support by the Ministry of Education and Science of the
Russian Federation Project No. 3.1463.2017, RFBR 15-03-07644 and RFBR 18-03-01220.
The computations were performed 
at the Resource Center "Computer Center of SPbU", St. Petersburg, Russia;
the Research Center for Computational Science, Okazaki, Japan;
the Research Institute for Information Technology at Kyushu University, Japan;
and the Numerical Materials Simulator, NIMS, Tsukuba, Japan.
We would like to thank Prof. Toshihiro Kondo, Prof. Takahisa Ohno and Dr. Nana Aoki for fruitful discussions
and Dr. Yuriko Ono for the help with visualisation of the crystal structures.
\end{acknowledgement}

%\bibliography{Au_catalysis}

\providecommand{\latin}[1]{#1}
\makeatletter
\providecommand{\doi}
  {\begingroup\let\do\@makeother\dospecials
  \catcode`\{=1 \catcode`\}=2 \doi@aux}
\providecommand{\doi@aux}[1]{\endgroup\texttt{#1}}
\makeatother
\providecommand*\mcitethebibliography{\thebibliography}
\csname @ifundefined\endcsname{endmcitethebibliography}
  {\let\endmcitethebibliography\endthebibliography}{}

\end{document}